\documentclass[]{IEEEtran}

\usepackage{amsmath}
\usepackage{multicol}
\usepackage{graphicx}
\usepackage[normalem]{ulem}
\usepackage{color}

\begin{document}

%
\title{ Accurate Closed-Form Real-Time EGN Model Formula Leveraging Machine-Learning over 8500 Thoroughly Randomized Full C-Band Systems}
%
%
%

\author{M.~Ranjbar~Zefreh, F.~Forghieri, S.~Piciaccia, P.~Poggiolini, 
\textit{Fellow,} OSA, \textit{Fellow,} IEEE
\thanks{M.~Ranjbar Zefreh and P.~Poggiolini are with Politecnico di Torino, Torino, Italy, e-mail: see www.optcom.polito.it. F.~Forghieri and S.~Piciaccia are with CISCO Photonics, Vimercate (MB), Italy.}
\thanks{ This study was supported by the CISCO Sponsored-Research Agreement (SRA) ``OptSys 2020'', and by the PhotoNext Center of Politecnico di Torino.
}
\thanks{Manuscript received December 17$^{\rm th}$, 2019.}}

\maketitle

\begin{abstract}
We derived an approximate non-linear interference (NLI) closed-form model (CFM), capable of handling a very broad range of optical WDM system scenarios. We tested the CFM over 8500  randomized C-band WDM systems, of which 6250 were fully-loaded and 2250 were partially loaded. The systems had highly diversified channel formats, symbol rates, fibers, as well as other parameters. We improved the CFM accuracy by augmenting the formula with simple machine-learning factors, optimized by leveraging the  system test-set. We further improved the CFM by adding a term which models special situations where NLI has high self-coherence. In the end, we obtained a very good match with the results found using the numerically-integrated Enhanced GN-model (or EGN-model).  We also checked the CFM accuracy by comparing its predictions with full-C-Band split-step simulations of 300 randomized systems. The combined high accuracy and very fast computation time (milliseconds) of the CFM potentially make it an effective tool for real-time physical-layer-aware optical network management and control.
\end{abstract}


%
\IEEEpeerreviewmaketitle

\section{Introduction}
%
%
%
%
\IEEEPARstart{P}{hysical}-layer-aware control and optimization of ultra-high-capacity optical networks is becoming an increasingly important aspect of networking, as throughput demand and loads increase. A necessary pre-requisite to achieve it, is the availability of accurate analytical modeling of fiber non-linear effects (or NLI, Non-Linear-Interference).  

Several NLI models have been proposed over the years, such as `time-domain' \cite{2012_JLT_Mecozzi, 2013_OE_Dar}, GN \cite{2014_JLT_Poggiolini}, EGN \cite{2013_OE_Dar, 2014_OE_Carena,2015_JLT_Serena}, as well as \cite{2011_OE_Bononi}-\cite{2016_JLT_Dar}, and several others, including various precursors of the former (see for instance refs. in \cite{2017_JLT_Poggiolini}). These NLI models, however, either contain integrals that make them unsuitable for real-time use, or otherwise assume too idealized system set-ups. The challenge is to derive approximate closed-form formulas, thus enabling real-time computation, that both preserve accuracy and are general enough to model highly diverse actual deployed systems.

In the GN/EGN model class, a rather general closed-form  set of formulas (or closed-form model, CFM) has been available for several years (Eqs. (41)-(43) in \cite{2014_JLT_Poggiolini}). These formulas, that we call CFM0, are a closed-form approximation of the incoherent GN-model (or iGN model \cite{2014_JLT_Poggiolini}).  They already allow to model systems with arbitrarily assorted WDM combs and non-identical spans and amplifiers. However, they do not support, among other things, dispersion slope and frequency-dependent loss, all-important features to enable the real-time modeling of actually deployed realistic systems and networks.
We upgraded CFM0 to include such missing features, following the approach proposed in \cite{2018_arXiv_Semrau}, \cite{2018_arXiv_Poggiolini}. We call these new formulas CFM1.


For a CFM   to be a viable candidate to support real-time physical-layer awareness, it must prove to be accurate over a wide variety of system configurations, ideally  spanning all practically possible general scenarios. We therefore set out to test CFM1 over a very large number (8500) of highly-randomized C-band WDM systems, of which 6250 were fully-loaded and the remainder (2250) partially-loaded. To the best of our knowledge, this is the first time such an extensive study has been performed.  The test consisted in comparing the  system  signal-to-noise ratio (SNR, inclusive of NLI), estimated using CFM1, with a \emph{benchmark}.  The benchmark we used is the full-fledged numerically-integrated EGN-model \cite{2014_OE_Carena}, which has  been shown to be very accurate in a  wide variety of system scenarios  \cite{2013_OE_Dar, 2014_OE_Carena, 2015_JLT_Serena, 2017_JLT_Poggiolini, 2018_OFC_Poggiolini, 2018_ECOC_Poggiolini}. 

The results of the comparison showed a reasonably good match with the EGN benchmark, overall, despite the many approximations involved in the derivation of CFM1 and the challenging  features of the system test-set.   However, CFM1 does suffer from an average tendency  towards \emph{underestimating} the SNR. This could be expected since CFM1 is derived from the GN-model, whose known behavior is to somewhat overestimate NLI \cite{2014_JLT_Poggiolini}.  In addition to such pessimistic bias, we also observed a substantial \emph{variance} of the error. 

To improve the accuracy of CFM1 vs. the EGN benchmark, we  leveraged the  system test-set to find a simple correction law which contains both  physical system parameters and  best-fitted coefficients, with the goal of turning CFM1 from a GN-model emulating CFM into an accurate \emph{EGN}-\emph{model} emulating CFM. This approach, that can be viewed as machine-learning over a big-data set,  proved  effective: the SNR estimation error of the new model, which we call CFM2, vs.~the EGN benchmark dropped dramatically.
Specifically, the bias towards underestimating SNR completely disappeared. The error variance reduced very substantially as well.

We noticed however a remaining problem of elevated SNR estimation error in a few outlier systems that, despite the small value of the error variance over the whole test-set, kept the \emph{peak} \emph{error} at a relatively high value. We examined the outlier systems features and found that they were characterized by   substantial \emph{NLI} \emph{coherence} effects. Since the CFMs (all versions) were originally based on the  \emph{incoherent} NLI\ accumulation assumption, high coherence in NLI causes large errors. We therefore  added to CFM2 a further term that approximates NLI\ coherence effects, following the approach reported in \cite{2019_arXiv_Poggiolini}. With the addition of such term, the new formulas, that we call CFM3, performed better on all accounts and, in particular, drastically curtailed peak error.
Overall, with the only exception of very-low dispersion set-ups operating at $D<2$ ps/(nm km), the very extensive testing showed CFM3 to be an effective and reliable approximation of the EGN model, across the very wide variety of links that the  8500 system test-set includes. A final refinement consisted in accounting for the effect of channel roll-off, resulting in CFM4, which was also tested over the  systems test-set.

As a further effort to confirm the validity and reliability of the approach, we compared the predictions of CFM4 with 300 \emph{full C-band split-step simulations} of systems randomly taken from the  system test-set. The results confirmed the previous conclusion.  

In the following, we first introduce CFM1 in the version \cite{2018_arXiv_Poggiolini}. Then, the features of the randomized  system large-test-set are described. Next, we show the accuracy results for CFM1, vs. the GN and  EGN models. Following, we introduce the machine-learning-based corrections aimed at improving the accuracy of CFM1 vs. the EGN model, obtaining CFM2. Then, we discuss the outlier cases and augment CFM2 with a NLI coherence-correction term, resulting in CFM3, which is then tested. CFM4, accounting for channel roll-off, is  introduced next and tested. Then, the accuracy of CFM4 vs. split-step simulations is assessed. Finally, computational effort is estimated.
Conclusions follow.

The CFM1 formulas in \cite{2018_arXiv_Poggiolini} can account for the impact of ISRS (Inter-Channel Stimulated Raman Scattering) on NLI generation, too. In this paper, however, ISRS is not considered. The testing of a version of CFM4 supporting ISRS is underway. For C-band systems, as considered here, this is a minor limitation.

A preliminary report on the research presented here was the subject of the ECOC 2019 paper \cite{2019_ECOC_Ranjbar}. Here we use a substantially larger and more diversified test-set and investigate other machine-learning correction formulas. The section on the comparison with 300 full-C-band split-step simulations is completely new. Much more detail than in \cite{2019_ECOC_Ranjbar} is provided throughout. In \cite{2019_PTL_Ranjbar} we reported on a earlier version of CFM2 which supported a much narrower range of systems: only QAM formats of high cardinality and no Gaussian constellations. In addition testing there was performed on about a third of the systems addressed here, in a much less diversified context. No testing on split-step simulations was shown in \cite{2019_PTL_Ranjbar} and neither CFM3 or CFM4 were available yet.   

\begin{figure}
\center
\includegraphics[width=\columnwidth]{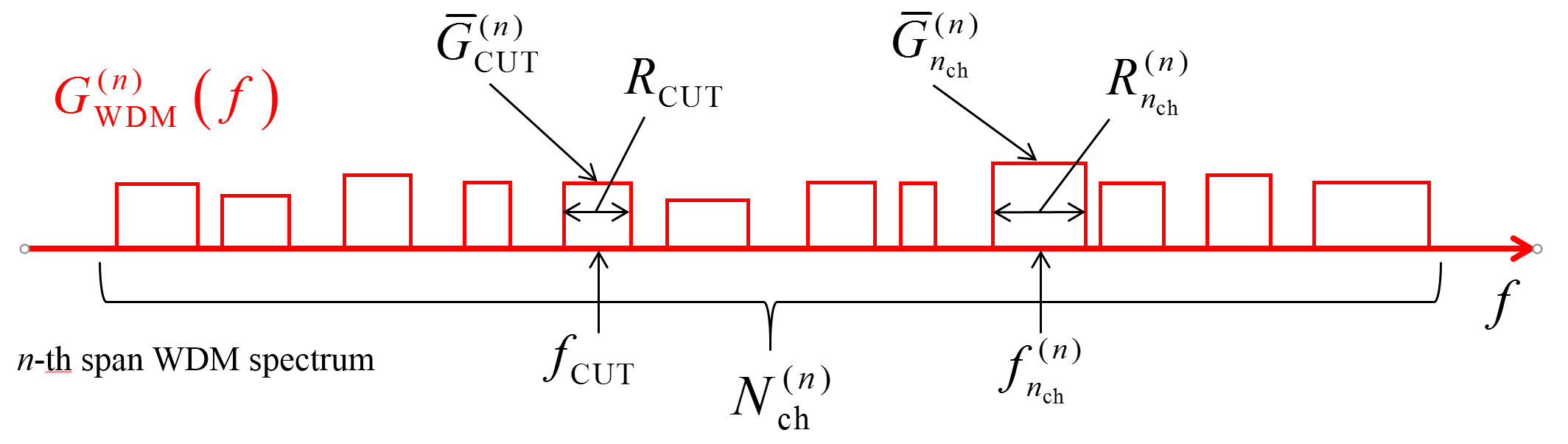}\vspace{0.3cm}
\includegraphics[width=7cm]{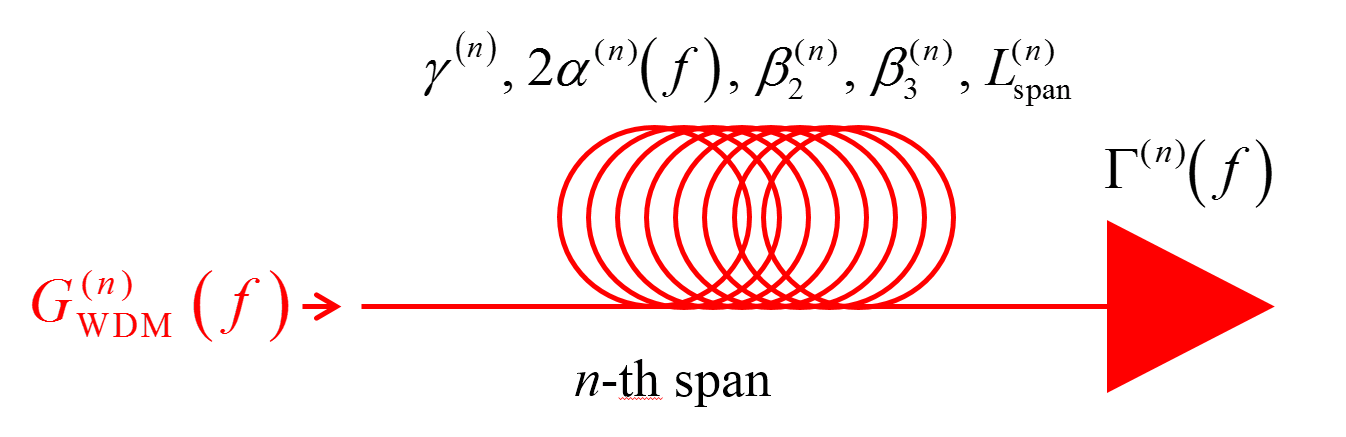}
\caption{
Top: the WDM comb PSD (power spectral density) $G_{_{\rm{WDM}}}^{({n})}\!\!\left( f \right)$ at the input of the $n$-th  span along the link. Bottom: the $n$-th  span, including fiber and the lumped elements transfer function $\,{\Gamma^{(n)}}\!\!\left( f \right)$. For the detailed definitions of all symbols, see Sect.~\ref{sect:CFM1}.
 \label{fig:layout}
 }\vspace{-0.4cm}
\end{figure}

\begin{figure*}[!t]
\normalsize
\begin{equation}
G_{_{\rm{NLI}}}^{{\rm{Rx}}}\left( {{f_{_{\rm{CUT}}}}} \right) = \sum\limits_{n = 1}^{{N_{{\rm{span}}}}} {\left( {G_{_{\rm{NLI}}}^{(n)}\left( {{f_{_{\rm{CUT}}}}} \right)\prod\limits_{k = n + 1}^{{N_{{\rm{span}}}}} {{\Gamma ^{(k)}}\!\!\left( {{f_{_{\rm{CUT}}}}} \right)\cdot{e^{ - 2 \cdot {\alpha ^{(k)}}\!\left( {{f_{_{\rm{CUT}}}}} \right) \cdot L_{{\rm{span}}}^{\left( k \right)}}}} } \right)} 
\label{eq:CFGN1}
\end{equation}

\begin{equation}
G_{_{\rm{NLI}}}^{(n)}\left( {{f_{_{\rm{CUT}}}}} \right) = \frac{{16}}{{27}}(\gamma ^{(n)})^2\,{\Gamma ^{(n)}}\!\!\left( {{f_{_{\rm{CUT}}}}} \right) \cdot {e^{ - 2{\alpha ^{(n)}}\!\left( {{f_{_{\rm{CUT}}}}} \right) \cdot \,L_{{\rm{span}}}^{(n)}}} \cdot \bar{G}_{_{\rm{CUT}}}^{(n)} \cdot \left( {{\rho^{(n)} _{_{{\rm{CUT}}}}} \cdot {{\left[ {\bar{G}_{_{\rm{CUT}}}^{(n)}} \right]}^2}I_{_{\rm{CUT}}}^{(n)} +\!\!\!\!\!\!\!\!\! \sum\limits_{{n_{{\rm{ch}}}} = 1\;,\;{n_{{\rm{ch}}}} \ne {n^{(n)}_{{\rm{CUT}}}}}^{{N^{(n)}_{{\rm{ch}}}}} \!\!\!\!\!\!\!\!\!\!\!\!\!{2{\rho^{(n)} _{{n_{{\rm{ch}}}}}} \cdot {{\left[ {\bar{G}_{{n_{{\rm{ch}}}}}^{(n)}} \right]}^2}} I_{{n_{{\rm{ch}}}}}^{(n)}} \right)
\label{eq:CFGN2}
\end{equation}


\begin{equation}
I_{_{\rm{CUT}}}^{(n)} = \frac{1}{{2\pi \mid \bar \beta _{2,_{\rm{CUT}}}^{\left( n \right)} \mid \cdot 2{\alpha ^{(n)}}\!\left( {{f_{_{\rm{CUT}}}}} \right)\;}}\,\cdot {{\rm{asinh}}\left( {\frac{{{\pi ^2}}}{2}\left| {\frac{{\bar \beta _{2,_{\rm{CUT}}}^{\left( n \right)}}}{{{2\alpha ^{(n)}}\!\left( {{f_{_{\rm{CUT}}}}} \right)}}} \right|R_{_{\rm{CUT}}}^2} \right)} 
\label{eq:ICUT}
\end{equation}


\begin{equation}
I_{{n_{{\rm{ch}}}}}^{(n)} = \frac{{{\rm{asinh}}\left( {{\pi ^2}\left| {\frac{{\bar \beta _{2,{n_{{\rm{ch}}}}}^{\left( n \right)}}}{{2{\alpha^{(n)}}\!\left( {f_{{n_{{\rm{ch}}}}}^{\left( n \right)}} \right)}}} \right|\left[ {f_{{n_{{\rm{ch}}}}}^{\left( n \right)} - {f_{_{\rm{CUT}}}} + \frac{{R_{{n_{{\rm{ch}}}}}^{\left( n \right)}}}{2}} \right]{R_{_{\rm{CUT}}}}} \right) - {\rm{asinh}}\left( {{\pi ^2}\left| {\frac{{\bar \beta _{2,{n_{{\rm{ch}}}}}^{\left( n \right)}}}{{2{\alpha^{(n)}}\!\left( {f_{{n_{{\rm{ch}}}}}^{\left( n \right)}} \right)}}} \right|\left[ {f_{{n_{{\rm{ch}}}}}^{\left( n \right)} - {f_{_{\rm{CUT}}}} - \frac{{R_{{n_{{\rm{ch}}}}}^{\left( n \right)}}}{2}} \right]{R_{_{\rm{CUT}}}}} \right)}}{{4\pi \left| {\bar \beta _{2,{n_{{\rm{ch}}}}}^{\left( n \right)}} \right| \cdot 2{\alpha ^{(n)}}\!\!\left( {f_{{n_{{\rm{ch}}}}}^{\left( n \right)}} \right)}}
\label{eq:IXCI}
\end{equation}

\begin{equation}\bar \beta _{2,_{\rm{CUT}}}^{\left( n \right)} = \beta _2^{(n)} + \pi \beta _3^{(n)}\left[ {2{f_{_{\rm{CUT}}}} - 2f_c^{(n)}} \right]\quad\quad ,\quad\quad \bar \beta _{2,{n_{{\rm{ch}}}}}^{\left( n \right)} = \beta _2^{(n)} + \pi \beta _3^{(n)}\left[ {f_{{n_{{\rm{ch}}}}}^{(n)} + {f_{_{\rm{CUT}}}} - 2f_c^{(n)}} \right]\label{eq:beta}
\end{equation}

\hrulefill
\vspace*{4pt}
\end{figure*}

\section{The Closed-Form Model 1}
\label{sect:CFM1}

The CFM1 formulas are Eqs.~(\ref{eq:CFGN1})-(\ref{eq:beta}), as derived in \cite{2018_arXiv_Poggiolini}. They are shown at the top of next page. 

As general notation remarks, all quantities bearing a superscript `$(n)$' or `$(k)$' are related  to the $n$-th or $k$-th \emph{span} in the link. Quantities referring to a specific channel bear the subscript $n_{\rm ch}$, which is an integer index that can span over all WDM channels. The subscript `CUT' identifies the \emph{channel under test}, i.e., the one whose performance is being estimated. When introducing physical quantities below, a coherent set of units is provided for the readers' convenience (other sets are of course possible). Fig.~1 is provided as a visual aid in the definition of various quantities related to the $n$-th span.

$G_{_{\rm{NLI}}}^{{\rm{Rx}}}\left({{f_{_{\rm{CUT}}}}}\right)$ in Eq.~(\ref{eq:CFGN1}) is the total power-spectral-density (PSD) of NLI at the receiver (Rx) and at the frequency $f_{_{\rm CUT}}$ of the channel under test. Frequencies and bandwidths are assumed to be expressed as THz and PSDs  as W/THz. 

Eq.~(\ref{eq:CFGN1}) shows 
$G_{_{\rm{NLI}}}^{{\rm{Rx}}}\left({{f_{_{\rm{CUT}}}}}\right)$  to be the sum over all spans of ${G_{_{\rm{NLI}}}^{(n)}}\left({{f_{_{\rm{CUT}}}}}\right)$. The latter is the PSD of NLI produced in the $n$-th span alone at $f_{_{\rm CUT}}$, assessed at the end of the span. The product operator `$\Pi$' in Eq. (\ref{eq:CFGN1}) accounts for the linear propagation of the PSD ${G_{_{\rm{NLI}}}^{(n)}}\left({{f_{_{\rm{CUT}}}}}\right)$ from the $n$-th span to the Rx.  

The other symbols in Eq.~(\ref{eq:CFGN1}), {\it all related to the $n$-th span}, are (see also Fig.~\ref{fig:layout}): $\Gamma_n\!\left(f \right)$, the power-gain/loss at frequency $f$ due to lumped elements, such as amplifiers and gain-flattening filters (GFFs), placed at the end of the span fiber; $ 2{\alpha _n}\!\left( f \right) $, the fiber power-loss coefficient (1/km) at frequency $f$; $L^{(n)}_{\rm span}$, the span length (km).

The  ${G_{_{\rm{NLI}}}^{(n)}}\left({{f_{_{\rm{CUT}}}}}\right)$ terms that feed  Eq.~(\ref{eq:CFGN1}) are found through Eq.~(\ref{eq:CFGN2}), where all quantities are related to     the $n$-th span. In it, $\gamma_{n}$ is the fiber non-linearity coefficient 1/(W$\cdot$km). $\bar{G}_{_{\rm CUT}}^{(n)}$ and $\bar{G}_{{ n_{\rm ch}}}^{(n)}$ are the \emph{effective PSDs} of the CUT and of the $n_{\rm ch}$-th WDM channel (see Fig.\ref{fig:layout}). They are defined as:

\begin{equation}
\begin{array}{c}
\bar{G}_{_{\rm CUT}}^{(n)}=P_{_{\rm CUT}}^{(n)}/R_{_{\rm CUT}}\\[4pt]
\bar{G}_{{ n_{\rm ch}}}^{(n)}=P_{{ n_{\rm ch}}}^{(n)}/R^{(n)}_{n_{\rm ch}}
\end{array}
\end{equation}
where $P_{_{\rm CUT}}^{(n)}$ and $P_{{ n_{\rm ch}}}^{(n)}$ are the launched power (W)  and $R_{_{\rm CUT}}$, $R^{(n)}_{n_{\rm ch}}$ are the symbol rates (TBaud), for the CUT and the $n_{\rm ch}$-th channels, respectively. 

The round bracket on the right of Eq.~(\ref{eq:CFGN2}) contains two terms. One includes the factor  $ I_{_{\rm{CUT}}}^{(n)}$ and accounts for NLI due to the self-channel interference (SCI) of the CUT onto itself. The other term, which includes the factors  $ I_{n_{\rm{ch}}}^{(n)}$, accounts for the cross-channel interference (XCI) of each WDM channel with the CUT. For a precise definition of SCI, XCI and MCI, see \cite{2012_JLT_Poggiolini}.
The summation runs over all the WDM channels indices $n_{\rm ch}=1,\ldots, N^{(n)}_{\rm ch}$, excluding the CUT index $n_{_{\rm CUT}}^{(n)}$. Note that the CFM1 formulas Eqs.~(\ref{eq:CFGN1})-(\ref{eq:beta})  allow for the WDM comb to be different at each span. This is why all channel-related parameters, including the total number of channels $ N^{(n)}_{\rm ch}$,  depend on the span index $n$. The only channel that is assumed to propagate across the whole link is the CUT.

The factors $ I_{_{\rm{CUT}}}^{(n)}$ and $ I_{n_{\rm{ch}}}^{(n)}$, Eqs.~(\ref{eq:ICUT}) and (\ref{eq:IXCI}), derive from closed-form approximate solutions of the GN-model integrals, as shown in \cite{2018_arXiv_Poggiolini}. They contain: the center frequency  $f_{_{\rm CUT}}$ and  $f^{(n)}_{n_{\rm ch}}$  (THz), and the \emph{effective} dispersion ${\bar \beta _{2,_{\rm{CUT}}}^{\left( n \right)}}$ and ${\bar \beta _{2,{n_{{\rm{ch}}}}}^{\left( n \right)}}$ (ps$^2$/km), of the CUT\ and of the $n_{\rm ch}$-th channel,  respectively. The effective dispersions are defined in Eq.~(\ref{eq:beta}), where $\beta _2^{(n)}$ and $\beta _3^{(n)}$ are the dispersion (ps$^2$/km)  and dispersion slope (ps$^3$/km), respectively, of the $n$-th span fiber. The frequency $f^{(n)}_c$ is where $\beta _2^{(n)}$ and $\beta _3^{(n)}$ are measured in the $n$-th span. Note that the `effective dispersions' of Eq.~(\ref{eq:beta}) originate from an approximation needed to obtain a closed-form formula, which amounts to considering dispersion different from channel to channel, but constant over each individual channel bandwidth   \cite{2018_arXiv_Poggiolini}. 

Finally, in Eq.~(\ref{eq:CFGN2}) the two factors $\rho^{(n)}_{_{\rm CUT}}$ and $\rho^{(n)}_{n_{\rm ch}}$  are  `machine-learning' functions meant to   turn  Eqs.~(\ref{eq:CFGN1})-(\ref{eq:beta}) from a CFM approximating the  \emph{GN}-model into one approximating the \emph{EGN}-model. For CFM1 they are set to 1 and unused. They will be discussed when introducing CFM2, in Sect.~\ref{sect:CFM2}.

\section{The System Test Set and the Test Procedure}
\label{sec:test-set}

The test-set consisted of  different C-band WDM systems. The C-band was considered to be extended over a 5-THz frequency range, with center frequency $f_{c}=193.8$ THz. For each of these systems, we focused on a single channel, which we called `channel-under-test' (CUT). The CUT could be either the \emph{lowest, center, or highest frequency} channel in the comb. 

A physical-layer awareness enabling tool, such as the CFMs addressed here, must be dependable over the widest range of possible systems. We therefore thoroughly diversified and randomized the generated test-set of WDM systems used for testing.
The test-set contained five different categories of systems, listed in the following. 
\begin{enumerate}
\item 3150 \emph{fully-loaded} C-band systems using PM-QAM formats of cardinality 16, 32, 64, 128 and 256;
\item 1250 \emph{partially-loaded} C-band systems using the same formats as in (1);
\item 2650 \emph{fully-loaded} C-band systems using PM-QAM formats of cardinality 16, 32, 64, 128 and 256 as well as PM-Gaussian formats;
\item 970 \emph{partially-loaded} C-band systems using the same formats as in (3);
\item 480 \emph{fully-loaded} C-band systems using PM-QAM formats of cardinality 4, 8, 16, 32, 64, 128 and 256 as well as PM-Gaussian formats.
\end{enumerate}

Each WDM comb was generated by randomly assigning to each individual channel one of the formats listed above. Regarding categories (1) and (2), any format was equally likely. To obtain partial loading in (2), a fully populated comb was generated and then each channel was turned on or off with probability 1/2. The average load was hence 50\%, with a wide spread of load values. Regarding categories (3) and (4), when generating a channel in the comb, a random choice was first taken between PM-QAM and PM-Gaussian, with 50\% probability. The CUT too could then be either a PM-Gaussian or a PM-QAM channel, with probability 1/2. As for category (5), WDM combs were generated similarly to category (3) but the range of possible QAM formats was extended by including PM-8QAM and PM-QPSK and the CUT was forced to be one of these two formats.  

Raised-cosine channel PSDs were assumed for all channels, with roll-off randomly chosen for each channel with a uniform distribution between 0.05 and 0.25. The symbol rate of each channel in the WDM comb was randomly selected among the following: 32, 64, 96 and 128 GBaud.  For 90\% of the systems, this resulted in an assigned spectral slot size of 43.5, 87.5, 131.25, 175 GHz, respectively. However, 10\% of the systems were generated as ultra-dense WDM ones, so that the null-to-null channel spectral separation was randomly selected between 5 and 20 GHz, irrespective of symbol rates.   

\begin{figure}
\includegraphics[width=\columnwidth]{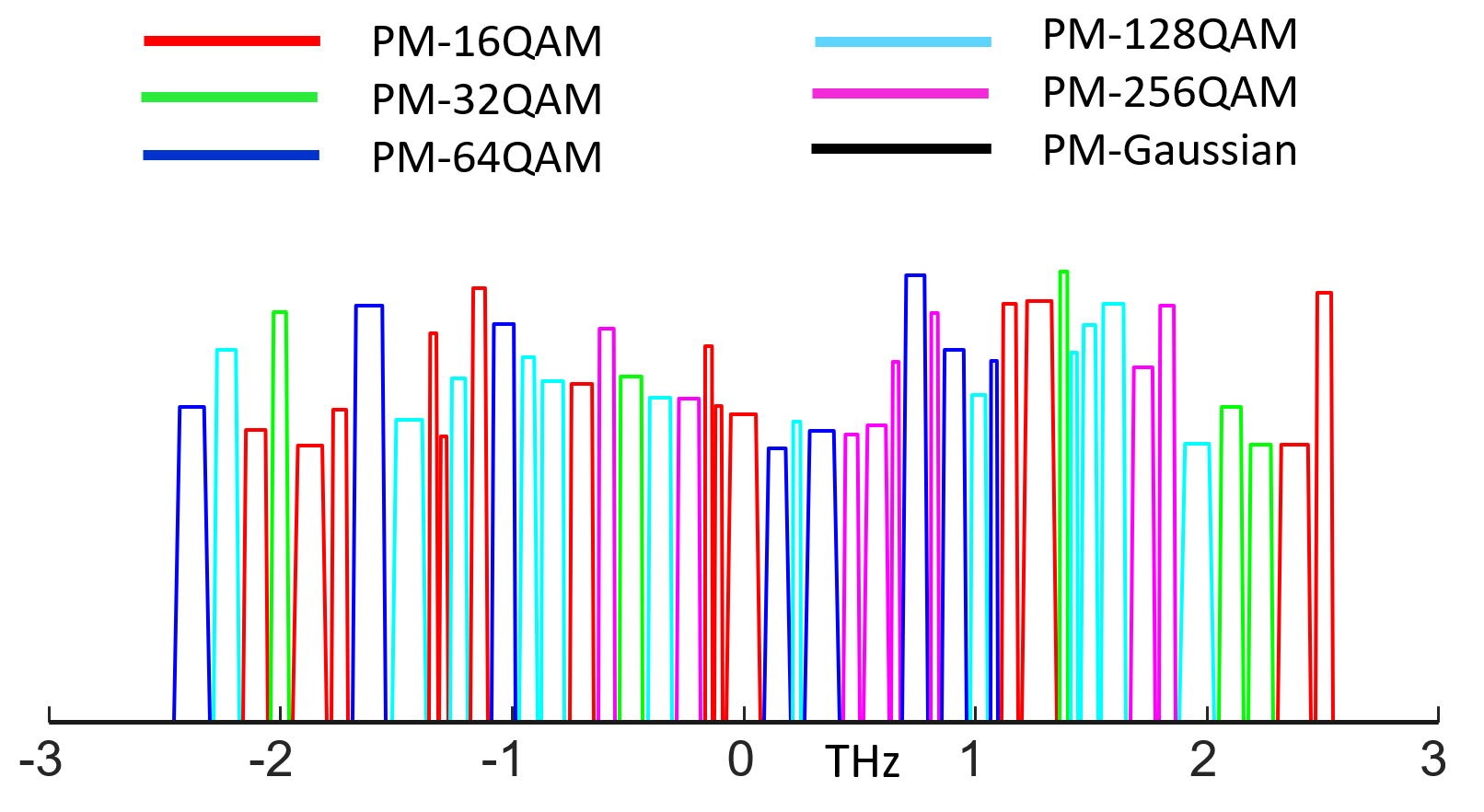}
\includegraphics[width=\columnwidth]{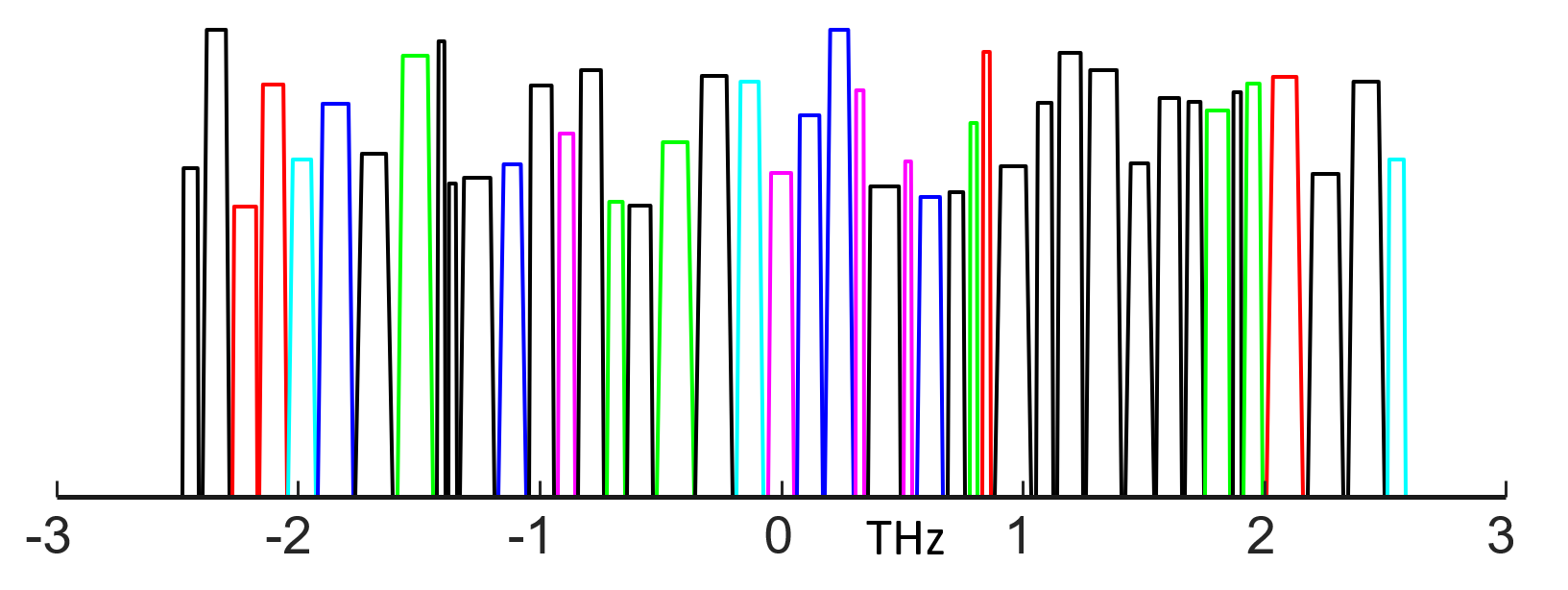}
\includegraphics[width=\columnwidth]{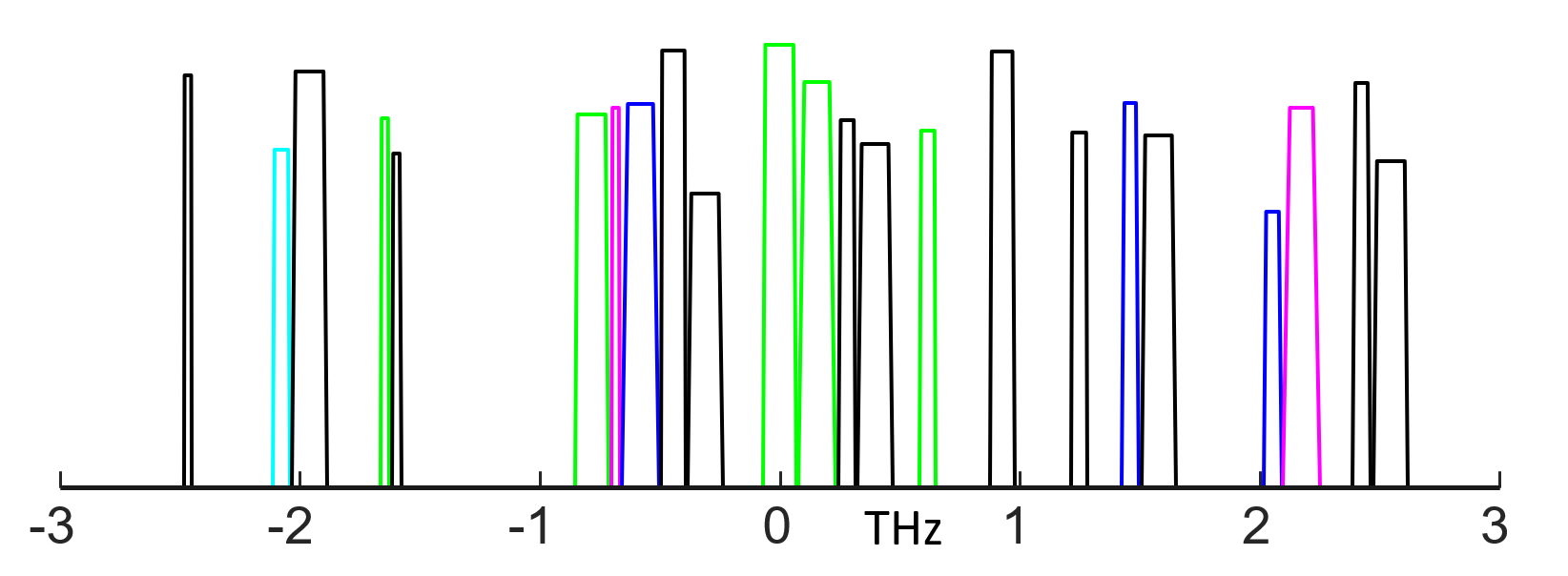}
\caption{ Three examples of WDM combs taken randomly from the 8500 systems test-set. Top: category 1. Middle: category 3. Bottom: category 4.
\label{fig:systems}}
\end{figure}

Fig.~\ref{fig:systems} shows three examples of generated WDM combs. From top to bottom, they are category (1), (3) and (4) respectively.

For each generated WDM comb, a different link was generated too.
Each link was made up of spans whose individual fiber was randomly chosen among three fiber types: SMF, NZDSF1 and NZDSF2. The fiber parameters were, respectively: loss, $\alpha_{\rm dB}$, 0.21, 0.22, 0.22 dB/km; dispersion at the WDM comb center (also assumed as center of the C-band, 193.8 THz), $\beta_{2}$, $-21.3$, $-4.85$, $-2.59$ ps$^2$/km; dispersion slope throughout the C-band, $\beta_{3}$, 0.1452, 0.1463, 0.1206   ps$^3$/km; non-linearity coefficient, $\gamma$, 1.3, 1.35, 1.77  (W$\cdot$km)$^{-1}$. Though not meaning to exactly reproduce any specific fiber, NZDSF1 and NZDSF2 have somewhat similar parameters to the commercial fibers E-LEAF$^{\rm TM}$ and TWC$^{\rm TM}$, respectively.
Each span length was generated randomly according to a uniform distribution between 80 and 120 km. Initially the test-set systems were generated with a fixed EDFA noise figure (NF). While accumulating more systems, we decided to add further realism by also randomizing this quantity. In the end, part of the systems have all EDFAs with NF 6 dB and the remainder with NF uniformly distributed between 5 and 6 dB. 

\begin{figure}
\includegraphics[width=\columnwidth]{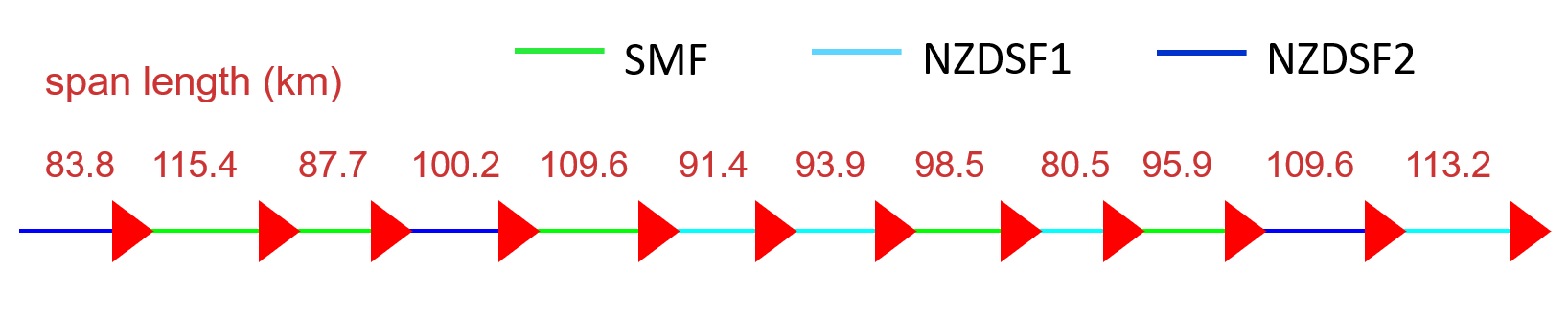}
\caption{ One example of the 8500 generated different test links.
\label{fig:links}}
\end{figure}

One example of the 8500 all-different generated links is shown in Fig.~\ref{fig:links}.

\subsection{The test procedure}
\label{sect:test_proc}
 Testing was based on assessing the accuracy of the estimate of the system  Signal-to-Noise Ratio (SNR), inclusive of NLI noise, defined as:
\begin{equation}{\rm{SN}}{{\rm{R}}} = \frac{{{P_{_{\rm{CUT}}}}}}{{{P_{_{\rm{ASE}}}} + {P_{_{\rm{NLI}}}}}}\label{eq:SNR}\end{equation}
where $P_{_{\rm ASE}}$ and $P_{_{\rm NLI}}$ are the noise powers affecting the received constellation, due to ASE and NLI, respectively, assuming a homodyne receiver with a matched filter.




In detail, the test procedure was as follows. For each system, SNR was first estimated using the closed-form model (CFM). We call such estimate ${\rm{SNR}}_{_{\rm{CFM}}}$. This  was then compared with another SNR estimate, found through a suitable {\em benchmark}, which we call ${\rm{SNR}}_{_{\rm{bmk}}}$. The benchmark  was either the GN-model, the EGN-model or split-step simulations.
The difference (in dB) between the two estimates constituted the  \emph{SNR estimation error vs. the benchmark}: 
\begin{equation}
\Delta^{{\rm{dB}}}_{_{\rm SNR}}={\rm{SNR}}_{_{\rm{CFM}}}^{{\rm{dB}}} - {\rm{SNR}}_{{\rm{bmk}}}^{{\rm{dB}}}
\label{eq:Delta}
\end{equation}

When calculating ${\rm{SNR}}_{_{\rm{CFM}}}$, the NLI\ power $P_{_{\rm NLI}}$ was approximated as follows: 
\begin{equation}
P_{\rm{NLI}} \approx  G_{\rm{NLI}}^{{\rm{Rx}}}
\left( {{f_{_{\rm{CUT}}}}} \right) \cdot R_{\rm {CUT}}
\label{eq:PNLICFM}
\end{equation}
where 
$G_{_{\rm{NLI}}}^{{\rm{Rx}}}\left( {{f_{_{\rm{CUT}}}}} \right)$ was estimated using a CFM. When calculating ${\rm{SNR}}_{_{\rm{bmk}}}$, the exact integral formula was used:
\begin{equation}
{P_{_{{\rm{NLI}}}}} = \int\limits_{ - \infty }^\infty  {G_{_{{\rm{NLI}}}}^{{\rm{Rx}}}\left( f+f_{_{\rm CUT}} \right)}  \cdot {\left| {H\left( f \right)} \right|^2}df
\label{eq:PNLIbmk}
\end{equation}
where $G_{_{\rm{NLI}}}^{{\rm{Rx}}}\left(f\right)$ was calculated using either the GN or EGN-model and  ${H\left( f \right)}$ was the receiver filter transfer function, matched to the root-raised-cosine spectral shape of the transmitted pulses, with the correct roll-off of the CUT. When the benchmark was a split step simulation (see Sect.~\ref{sect:split-step}), ${\rm{SNR}}_{_{\rm{bmk}}}$ was measured directly on the received constellation.

Two aspects are key in this procedure: where along the system was the error estimated and what launch power was assumed.
As for the former, the  SNR comparison between CFM and benchmark was carried out at the span number corresponding to the \emph{max-reach} for the CUT. The max-reach was accurately found using the EGN-model, based on the following SNR sensitivities: when  a   PM-QAM channel was the CUT, the assumed sensitivity values were: 5.18, 9.30, 11.48, 14.45, 17.00, 19.71, 22.33 dB, from constellation cardinality 4 to 256 in powers of two, respectively. These  SNR values correspond to a normalized generalized mutual information (NGMI) value of 0.87, for all formats.
They in turn correspond to pre-FEC BERs between  3.4$\cdot 10^{-2}$ and 3.7$\cdot 10^{-2}$, values that can be coped with using modern soft-decoding FECs. 

When the CUT was a  PM-Gaussian channel, such as it could occur for system categories (3) and (4), its SNR  sensitivity was set at follows. First, a mutual information (MI) target was randomly selected, between 6.96 and 13.92 b/s, with a uniform distribution. Then, using Shannon's law, the  SNR corresponding to the random MI target was found and used as SNR sensitivity for the PM-Gaussian channel. Note that the extremes of the uniform MI distribution correspond to the GMI values assumed for PM-16QAM (6.96 b/s) and PM-256QAM (13.92 b/s), respectively.

With regards to the launch power used for testing, a detailed explanation of the power optimization procedure is reported in Appendix \ref{app:powopt}. In brief, the WDM channels were launched, on average, at an approximately optimal power into each span  but, for further realism, we applied a random launch power deviation from optimum, uniformly distributed between  $\pm$30\%, different for each channel in the comb.  The CUT was instead launched at its optimal power without any random power deviation.

As a concluding remark to this section, the thorough randomization of combs and links led to a spread of system scenarios which was rather extreme. To mention one indicator, maximum reaches ranged from 1 span to 35 spans, essentially covering the overall practical range of terrestrial networks. This was done on purpose, to subject the CFMs to a very demanding `stress-test'.


\section{Testing CFM1}
\label{sec:CFM1}

\begin{figure}
\center
\includegraphics[width=\columnwidth]{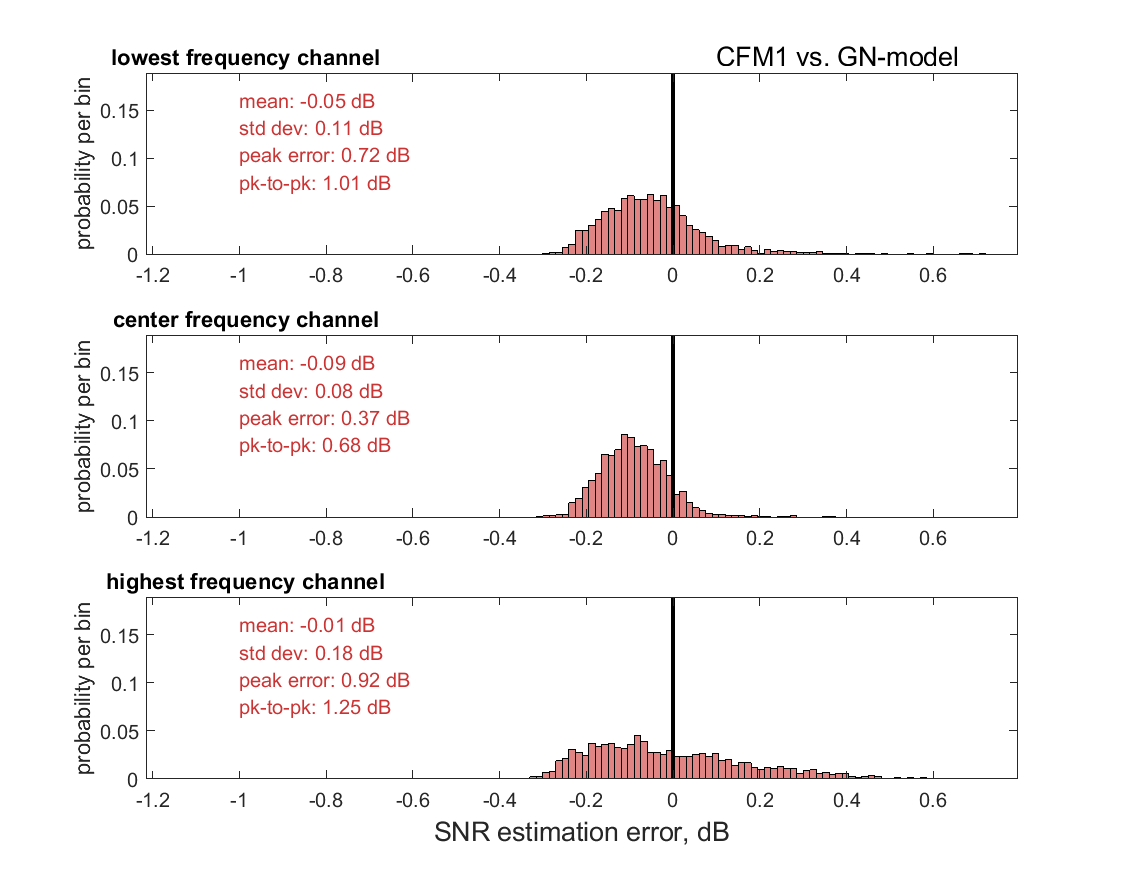}
\vspace{-0.8cm}
\caption{ Histograms of the SNR estimation error $\Delta^{{\rm{dB}}}_{_{\rm SNR}}$  between the closed-form model 1 (CFM1) and the GN-model, as defined by Eq.~(\ref{eq:Delta}). The error is measured at maximum reach.  Each  histogram was built by looking at  about 2800 different systems. 
\label{fig:CFM1-GN}}
\vspace{-0.4cm}
\end{figure}

Fig.~\ref{fig:CFM1-GN} shows the histograms of the SNR estimation error $\Delta^{{\rm{dB}}}_{_{\rm SNR}}$, as defined by Eq.~(\ref{eq:Delta}), between CFM1 and  the GN-model.  
The three plots, from top to bottom, address the case of the CUT\ being the lowest-frequency, center-frequency and highest frequency channels in the WDM comb.
Together with the histograms, the related\ mean, standard deviation  $\sigma$, peak and peak-to-peak error are displayed.

The histograms show a  small mean error of less than $-0.1$~dB on all three CUTs, The standard deviation is also rather small, especially for the low and center frequency CUTs. Peak absolute error was however rather large, signaling the presence of outliers. Overall, though, we considered these initial results  rather encouraging, given the extent of the approximations used to obtain CFM1 and the extreme diversity of systems addressed in the test-set. Some of the discrepancy can also be attributed to the fact that CFM1 is an approximation of the \emph{incoherent} GN-model, whereas the benchmark used in Fig.~\ref{fig:CFM1-GN} was the standard (coherent) GN-model (see \cite{2014_JLT_Poggiolini}, \cite{2012_JLT_Poggiolini}, for the difference between the two).

\begin{figure}
\center
\includegraphics[width=\columnwidth]{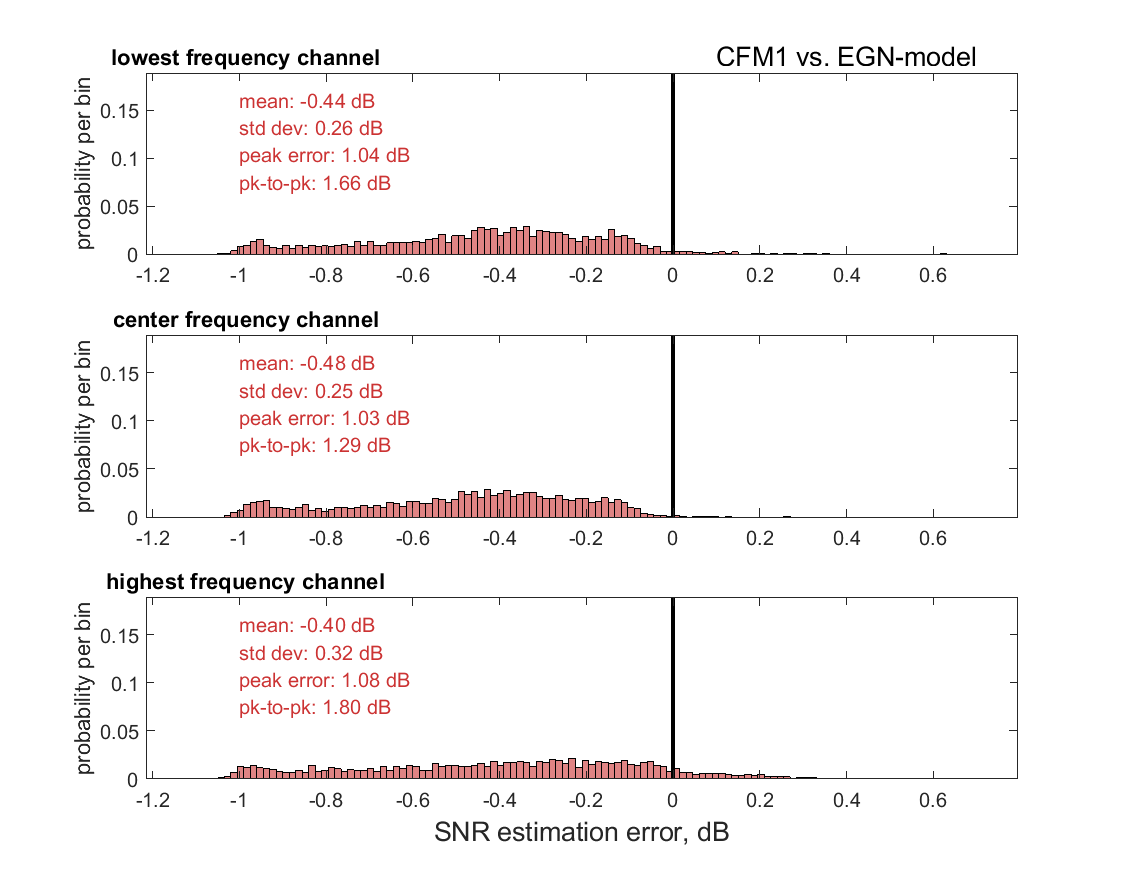}
\vspace{-0.8cm}
\caption{  Histograms of the SNR estimation error  $\Delta^{{\rm{dB}}}_{_{\rm SNR}}$ between the closed-form model 1 (CFM1) and the EGN-model, as defined by Eq.~(\ref{eq:Delta}). The error is measured at maximum reach. Each  histogram was built by looking at about 2800 different systems.   
\label{fig:CFM1-EGN}}
\vspace{-0.4cm}
\end{figure}

We then proceeded to compare CFM1 with the numerically-integrated EGN-model, complete with all its terms (see \cite{2014_OE_Carena}). The    SNR estimation error $\Delta^{{\rm{dB}}}_{_{\rm SNR}}$  results are displayed in Fig.~\ref{fig:CFM1-EGN}. The plots clearly show that the performance of CFM1 is much less favorable when compared to the EGN-model. The   reason is that CFM1 was derived from the GN-model and not from the EGN. Quite telling in this regard is the mean value error which is now about -0.45 dB and is attributable to the known skew between GN and EGN-model, with the GN being somewhat pessimistic. Apart from mean value, both the standard deviation and peak absolute error increase substantially. 

Since the goal for the CFM is that of being a reliable and accurate practical modeling tool, then it is the EGN model that must be used as main benchmark, being the more accurate between GN and EGN. We therefore proceeded to look for suitable strategies to turn CFM1 into a more faithful approximation of the EGN-model.     

\section{CFM2 derivation and test}
\label{sect:CFM2}
To improve the accuracy of CFM1 vs. the EGN-model, we decided to leverage the large system test-set and use a machine-learning strategy to fine-tune suitable `correction factors' that could turn  Eqs.~(\ref{eq:CFGN1})-(\ref{eq:beta}) into a better approximation of the \emph{EGN}-model.  

The machine-learning factors are  $\rho^{(n)} _{_{{\rm{CUT}}}}$
and  $\rho^{(n)} _{n_{{\rm{ch}}}}$ in Eq.~(\ref{eq:CFGN2}), which for CFM1 were not used (they were  set to 1). To obtain the improved closed-form model CFM2 we defined them as:
\begin{equation}
\begin{array}{l}
\rho _{{n_{{\rm{ch}}}}}^{(n)} = {a_1} + {a_2} \cdot {\Phi _{{n_{{\rm{ch}}}}}^{a_3}} + {a_4} \cdot {\Phi _{{n_{{\rm{ch}}}}}^{a_5}} \cdot  \\ \hspace{3 cm} {\left( 1+ {a_6} \cdot  \left[\,{\left| {{\beta _{2,{\rm{acc}}}}\left( {n,{n_{{\rm{ch}}}}} \right)}  \right|  }+{a_{7}}\,\right]^{{a_{8}}} \right)} \vspace{0.2 cm}
\\ 
\rho _{_{{\rm{CUT}}}}^{\left( n \right)} ={a_{9}} + {a_{10}} \cdot {\Phi _{_{\rm CUT}}^{a_{11}}} + {a_{12}} \cdot {\Phi _{_{\rm CUT}}^{a_{13}}} \cdot  \\ \hspace{1 cm} {\left( 1+ {a_{14}} \cdot R_{_{\rm CUT}}^{a_{15}} + {a_{16}} \cdot  \left[\,{\left| {{\beta _{2,{\rm{acc}}}}\left( {n,{n_{_{\rm{CUT}}}}} \right)}  \right|  }+{a_{17}}\right]^{{a_{18}}} \right)}
\\
\end{array}
\label{eq:rho2}
\end{equation}
\noindent where $a_{1}\dots a_{18}$ are free parameters that need to be optimized and ${{\bar \beta }_{{\rm{2}}{\rm{,acc}}}}\left( {n,{n_{{\rm{ch}}}}} \right)$ is the effective accumulated dispersion at the $n_{\rm ch}$-th channel frequency, from link start till the input of the $n$-th span fiber, defined as:
\begin{equation*}
{{\bar \beta }_{{\rm{2}},{\rm{acc}}}}\left( {n,{n_{{\rm{ch}}}}} \right) = \sum\limits_{k = 1}^{n - 1} {\bar \beta _{2,{n_{{\rm{ch}}}}}^{\left( k \right)} \cdot L_{{\rm{span}}}^{(k)}} 
\end{equation*}
The parameters $\Phi_{n_{\rm ch}}$ and $\Phi_{_{\rm CUT}}$ are the  constant $\Phi$ used by the EGN model, which depends on the modulation format of a  channel \cite{2014_OE_Carena}, \cite{2016_book_Poggiolini}. The exact values are shown in Table \ref{tab:Phi}, for the modulation formats used in this paper and for PM-BPSK. The addressed channels are either that of the generic $n_{\rm ch}$-th channel (for $\Phi_{n_{\rm ch}}$) or of the CUT (for $\Phi_{_{\rm CUT}}$). 

\begin{table}
\centering
\caption{Exact values of the $\Phi$  parameter.
\label{tab:Phi}}
\begin{tabular}{|c|c|}\hline
format & $\Phi$  \\ \hline
PM-BPSK & 1  \\\hline
PM-QPSK & 1  \\\hline
PM-8QAM & 2/3  \\\hline
PM-16QAM & 17/25  \\\hline
PM-32QPSK & 69/100  \\\hline
PM-64QAM & 13/21  \\\hline
PM-128 & 1105/1681  \\\hline
PM-256 & 257/425  \\\hline
PM-Gaussian & 0 \\\hline
\end{tabular}
\end{table}

Eq.~(\ref{eq:rho2}) was conceived based on clues from \cite{2012_JLT_Poggiolini} and \cite{2015_JLT_Poggiolini} and on an extensive numerical study of NLI  estimation error sensitivity vs. various system \textit{physical} parameters, which turned out to favor $R_{_{\rm CUT}}$ and  ${{\bar \beta }_{{\rm{2}}{\rm{,acc}}}}$ as the most effective ones. For more details on how this was done, please see Appendix \ref{app:corr}. See also Appendix \ref{app:sens} for an investigation of NLI estimation error dependence  on the choice of the physical parameters in the machine-learning factors and for  comments on its specific analytical layout. Note that in Sect.~\ref{sect:roll-off} we will introduce an alternate form of  $\rho^{(n)} _{_{{\rm{CUT}}}}$
and  $\rho^{(n)} _{n_{{\rm{ch}}}}$ which takes channel roll-off into account as well. In this section, we make use of Eq.~(\ref{eq:rho2}).

The free parameters  $a_{1}$-$a_{18}$ were optimized using a `machine-learning' approach, as follows. Out of the 8500 systems of the test-set, 1500 were selected as `training' set. Of them, 750 were fully-loaded and 750 were sparsely-loaded. The `cost-function'  to be minimized was the sum of error contributions of the form:
\begin{equation}
\Delta_{\rm NLI}= \frac{\mid P_{_{\rm{NLI,CFM}}}^{(n)}-P_{_{\rm{NLI, EGN}}}^{(n)}\mid^2 }{\mid P_{_{\rm{NLI, EGN}}}^{(n)}\mid^2 }
\label{eq:Delta_NLI}
\end{equation}
where $P_{_{\rm{NLI,CFM}}}^{(n)}$ is the power of NLI noise affecting the CUT constellation, assessed at the $n$-th span of a system in the training-set, estimated using CFM2 according to Eq.~(\ref{eq:PNLICFM}), and $ P_{_{\rm{NLI, EGN}}}^{(n)}$ is the same quantity estimated through the EGN model, according to Eq.~(\ref{eq:PNLIbmk}).
$\Delta_{\rm NLI}$ was calculated \textit{at each span} of each system, from the first span to max-reach. As a result, despite using only 1500 systems for the training set, the cost-function was made up of about 11,500 error contributions of the form  Eq.~(\ref{eq:Delta_NLI}). The resulting optimized parameters are reported in Table \ref{tab:a}.
\begin{table}
\centering
\caption{Optimized values of the parameters $a_{1}$-$a_{18}$ for the machine-learning factors of CFM2 \label{tab:a}}
\begin{tabular}{||c|c||c|c||}\hline
parameter & value & parameter & value   \\ \hline
$a_{1}$ & +9.3143e-1 & $a_{10}$& -1.8838e0  \\\hline
$a_{2}$ & -7.7122e-1 & $a_{11}$& +6.2974e-1   \\\hline
$a_{3}$ & +9.1090e-1 & $a_{12}$&  -1.1421e+1  \\\hline
$a_{4}$ & -1.4555e+1 & $a_{13}$&  +6.7368e-1  \\\hline
$a_{5}$ & +8.5816e-1 & $a_{14}$& -1.1759e0   \\\hline
$a_{6}$ & -9.9415e-1 & $a_{15}$& +6.4482e-3   \\\hline
$a_{7}$ & +1.0812e0 & $a_{16}$&  +1.8738e+5  \\\hline
$a_{8}$ & +5.2247e-3 & $a_{17}$&   +1.9527e+3 \\\hline
$a_{9}$ & +9.9313e-1 & $a_{18}$& -2.0016e0   \\\hline

\end{tabular}
\end{table}

\begin{figure}
\center
\includegraphics[width=\columnwidth]{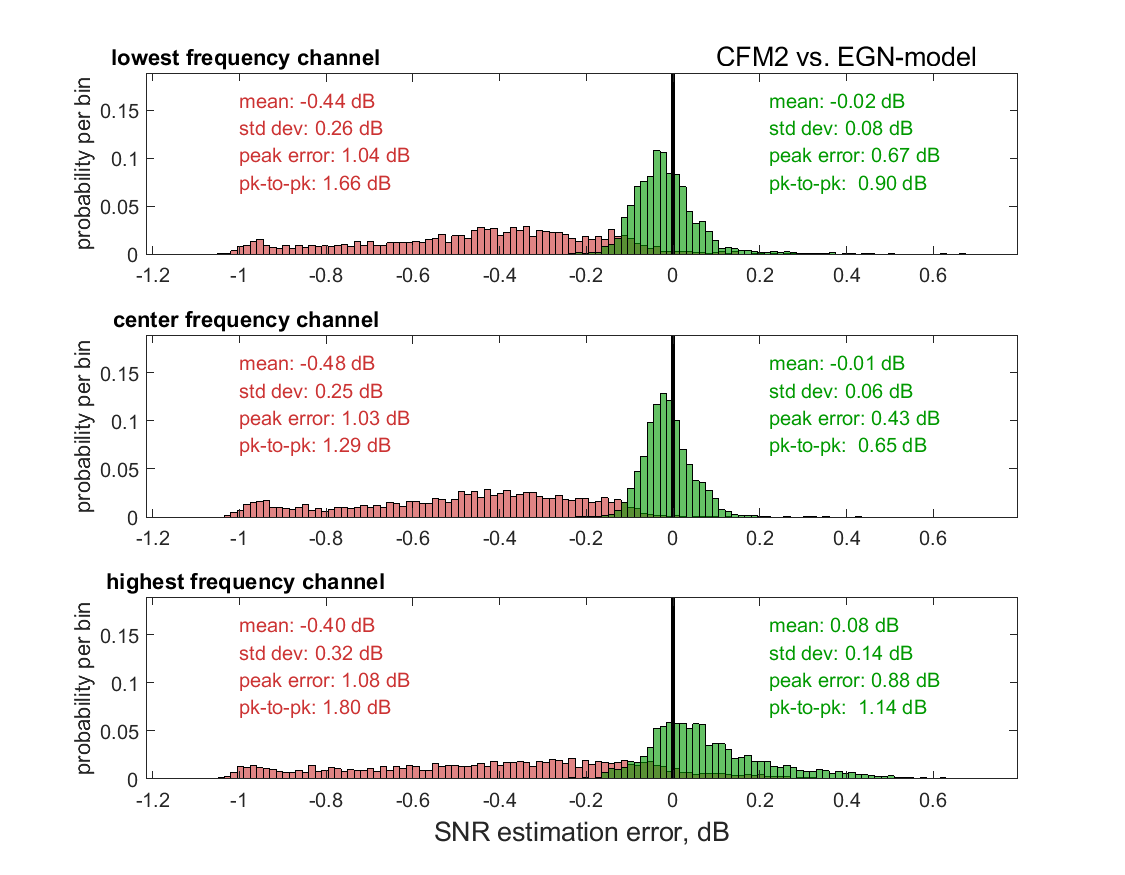}
\vspace{-0.4cm}
\caption{  Green bars: histograms of the SNR estimation error $\Delta^{{\rm{dB}}}_{_{\rm SNR}}$  between the closed-form model 2 (CFM2) and the EGN-model as defined by Eq.~(\ref{eq:Delta}). The error is measured at maximum reach. Each  histogram was built by looking at about 2800 different systems. The red bars are CFM1 vs. EGN-model, same as Fig.~\ref{fig:CFM1-EGN}, shown here for comparison.
\label{fig:CFM2-EGN}}
\end{figure}

We then ran   on CFM2 the same test procedure as described in Sect.~\ref{sect:test_proc}, over the complete 8500 system test-set, using the machine-learning factors Eq.~(\ref{eq:rho2}), with the optimized parameters of Table \ref{tab:a}. The new histograms are shown as green bars in Fig.~\ref{fig:CFM2-EGN}.  The histograms of CFM1 are also shown for comparison as red bars. The improvement is quite dramatic, both visually and numerically, as proved by the statistical parameters reported in the plots. The mean error is virtually zero, for the lowest and center frequency channel, and less than 0.1 dB for the high frequency channel. Standard deviations are very low. Overall, CFM2 is a much better model than CFM1 and already delivers quite convincing performance.

\begin{figure}
\center
\includegraphics[width=\columnwidth]{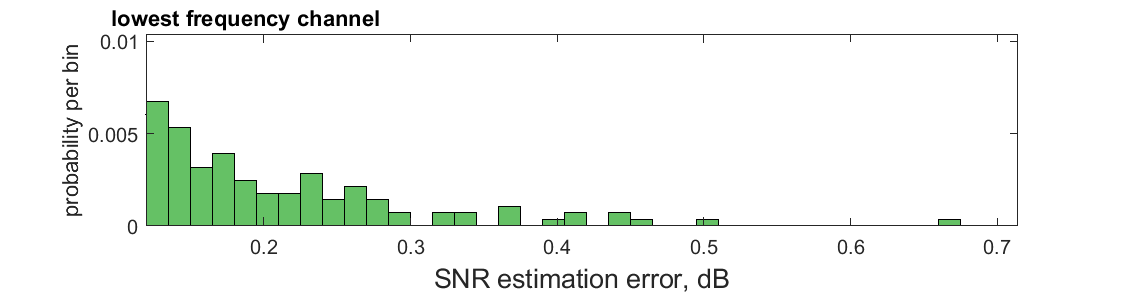}
\vspace{-0.4cm}
\caption{  Zoomed-in version of the lowest-frequency channel histogram from Fig.~\ref{fig:CFM2-EGN}, showing the farther outliers in the plot. Visible to the right is the single farthest outlier system located at $+0.67$~dB.   
\label{fig:blow-up}}
\end{figure}

This said, two aspects emerging from the plots and  the statistical indicators in Fig.~\ref{fig:CFM2-EGN} are not satisfactory. First of all, the high-frequency channels histogram is clearly more spread out than the other two and it is important to understand why. In addition, even though the error standard deviations are all small, the channels suffer from the presence of  outliers whose SNR estimation error is quite significant, as shown by the peak and peak-to-peak error indicators. Fig.~\ref{fig:blow-up} shows a blown-up section of the lowest frequency channel histogram from Fig.~\ref{fig:CFM2-EGN}, which displays some of the farthest outliers, almost invisible in Fig.~\ref{fig:CFM2-EGN}. Even though it is a relatively few cases, they indicate a weakness in the model which would be important to identify and, if possible, remove. In the following we first discuss the less favorable  high-frequency histogram. In Sect.~\ref{sect:CFM3} we introduce an improved closed-form model,  CFM3, to deal with the outliers.

\subsection{The high frequency CUTs}
We examined the system parameters of those highest-frequency CUTs that had a high SNR estimation error. We found that they had a substantial prevalence of NZDSF2 in the link, especially at the link start. The potential problem related to the presence of NZDSF2 is that such fiber has a very low dispersion value where the   highest-frequency CUTs are located. The value is  $\beta_{2}=-0.73$ ps$^2$/km, or $D=0.56$ ps/(nm$\cdot$km). At such low dispersion, some of the approximations used to derive the CFM formulas lose accuracy. Therefore, we tentatively attributed the less favorable performance of CFM2 for the highest-frequency CUTs to dispersion being too low for the model to handle. 

\begin{figure}
\center
\includegraphics[width=\columnwidth]{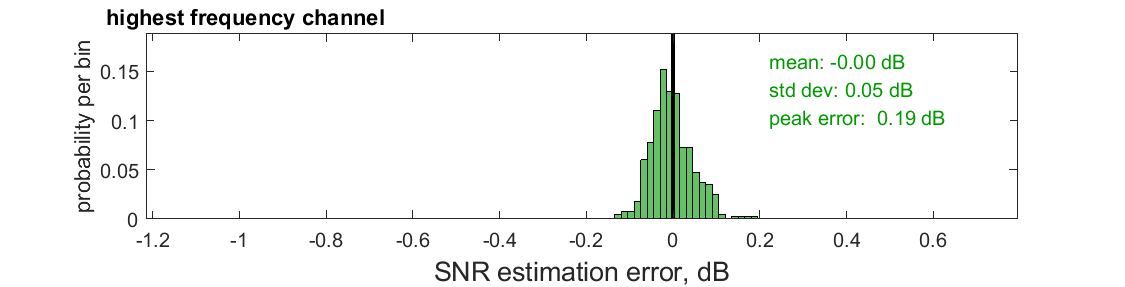}
\vspace{-0.4cm}
\caption{  Highest-frequency channels histograms of the SNR estimation error $\Delta^{{\rm{dB}}}_{_{\rm SNR}}$  between the closed-form model 2 (CFM2) and the EGN-model, as defined by Eq.~(\ref{eq:Delta}). The test-set for this plot was comprised of 400 systems not using fiber type  NZDSF2.    
\label{fig:no_NZDSF2}}
\end{figure}

We checked this hypothesis by generating a separate 400-system  test-set  using category (1) WDM combs, where only SMF and NZDSF1 were present, but no NZDFS2. Note that NZDSF1 too has its minimum dispersion (over the C-Band) where the   highest-frequency CUTs are located, but such minimum is   $\beta_{2}=-2.61$ ps$^2$/km, or $D=2.07$ ps/(nm$\cdot$km), that is, significantly higher than NZDSF2.
The histogram of the SNR estimation error $\Delta^{{\rm{dB}}}_{_{\rm SNR}}$ for the highest-frequency CUTs over this special test-set is shown in Fig.~\ref{fig:no_NZDSF2}.
The histogram now clearly appears  similar to those for the center and lowest-frequency channels shown in Fig.~\ref{fig:CFM2-EGN}. In particular, the standard deviation is comparably low. This evidence appears to confirm that it was the very low dispersion of NZDSF2 that generated a much larger histogram spread for the highest-frequency channel.

Based on these results, we do not recommend using the CFMs presented in this paper below about $|\beta_{2}| =2.5$ ps$^2$/km, or about $|D|=2$ ps/(nm$\cdot$km). To go lower, the CFM should be analytically modified, which we leave for future investigation.

\section{CFM3: combating outliers}
\label{sect:CFM3}
The presence of outliers in the error histograms cannot be ascribed to low dispersion. In particular, the lowest-frequency CUTs experience a dispersion    $\beta_{2}=-4.45$ ps$^2$/km, or $D=3.44$ ps/(nm$\cdot$km), which is sufficiently high to be dealt with by the models, as also previous investigations addressing simpler CFMs have indicated (see for instance \cite{2017_JLT_Poggiolini}, Sect.~IIE). Yet, Fig.~\ref{fig:blow-up} clearly shows that even the lowest-frequency CUTs histogram presents a substantial number of outliers. 

To find out what caused such outliers we again focused on the system details. Particularly telling is the outermost outlier in Fig.~\ref{fig:blow-up}, giving rise to a single-system bin placed a $\Delta^{{\rm{dB}}}_{_{\rm SNR}}=0.67$~dB. The system is the one shown in Fig.~\ref{fig:systems}, bottom, i.e., a category 4 (partial load, PM-QAM and PM-Gaussian). The CUT consists of a thin-looking 32~GBaud PM-Gaussian channel placed at the low-frequency edge of the C-band. The next channels happen by chance to be far, the nearest being located about 400 GHz away. Other big voids are present in the first 1.7 THz  from the CUT. Also, the link was long: max reach was 30 spans, since the PM-Gaussian CUT was in this instance assigned a SNR sensitivity of 10.3 dB. Many spans  (15) consisted of highly dispersive SMF. In these conditions, two things happened: 2/3 (in power) of the NLI noise was SCI (single-channel interference, i.e., NLI noise generated by the channel onto itself); such SCI noise was highly self-coherent. As a result, by far most of the overall NLI experienced by this CUT was highly self-coherent. 

The study of this and other outliers strongly suggested that it was indeed high NLI accumulation coherence that caused the large errors. High coherence is not handled well because the starting model for the derivation of all of CFMs 0, 1 and 2 is the \emph{incoherent} GN-model, whose founding approximation is incoherent NLI accumulation. Such models cannot be expected to accurately model situations where coherence is strong. 


To remedy this problem, a further closed-form term meant to approximately account for coherent accumulation of SCI was derived and added to  Eq.~(\ref{eq:ICUT}), i.e., the equation that specifically estimates SCI. The theory and the approximations used to obtain the extra term are reported in detail in \cite{2019_arXiv_Poggiolini}. As a result, the modified Eq.~(\ref{eq:ICUT}) becomes: 
\begin{equation}
\begin{array}{l}
I_{_{{\rm{CUT}}}}^{(n)} = \frac{1}{{2\pi \left| \bar \beta _{2{,_{{\rm{CUT}}}}}^{\left( n \right)} \right| \cdot 2{\alpha _n}\left( {{f_{_{{\rm{CUT}}}}}} \right)\;}}\left\{ {{\rm{asinh}}\!\left( {\frac{{{\pi ^2}}}{4}\left| {\frac{{\bar \beta _{2{,_{{\rm{CUT}}}}}^{\left( n \right)}}}{{{\alpha _n}\left( {{f_{_{{\rm{CUT}}}}}} \right)}}} \right|B_{_{\rm{CUT}}}^2} \right)  } \right.\\[12pt]
\left. { + 2\frac{{{\rm{Si}}\left( {{\pi ^2}\left|\bar \beta _{2{,_{{\rm{CUT}}}}}^{\left( n \right)}\right| L_{{\rm{span}}}^{(n)}B_{_{{\rm{CUT}}}}^2} \right)}}{{\pi {\kern 1pt} {\alpha _n}\left( {{f_{_{{\rm{CUT}}}}}} \right)L_{{\rm{span}}}^{(n)}\;}}\left[ {{\mathop{\rm HN}\nolimits} \left( {{N_{{\rm{span}}}} - 1} \right) + \frac{{1 - {N_{{\rm{span}}}}}}{{{N_{{\rm{span}}}}}}} \right]} \right\}
\end{array}\end{equation} 
where the added term is the one appearing on the bottom line. `HN' stands for harmonic number and `Si' for the sine-integral function.  This change in Eq.~(\ref{eq:ICUT}), together with using the correction factors Eq.~(\ref{eq:rho2}), give rise to what we call CFM3. Since the analytical make-up of the model was changed, even though the machine-learning factors still had the same definition Eq.~(\ref{eq:rho2}), we re-ran the optimization of the parameters $a_{1}$-$a_{18}$. The new values are shown in Table~\ref{tab:CFM3}.

\begin{figure}
\center
\includegraphics[width=\columnwidth]{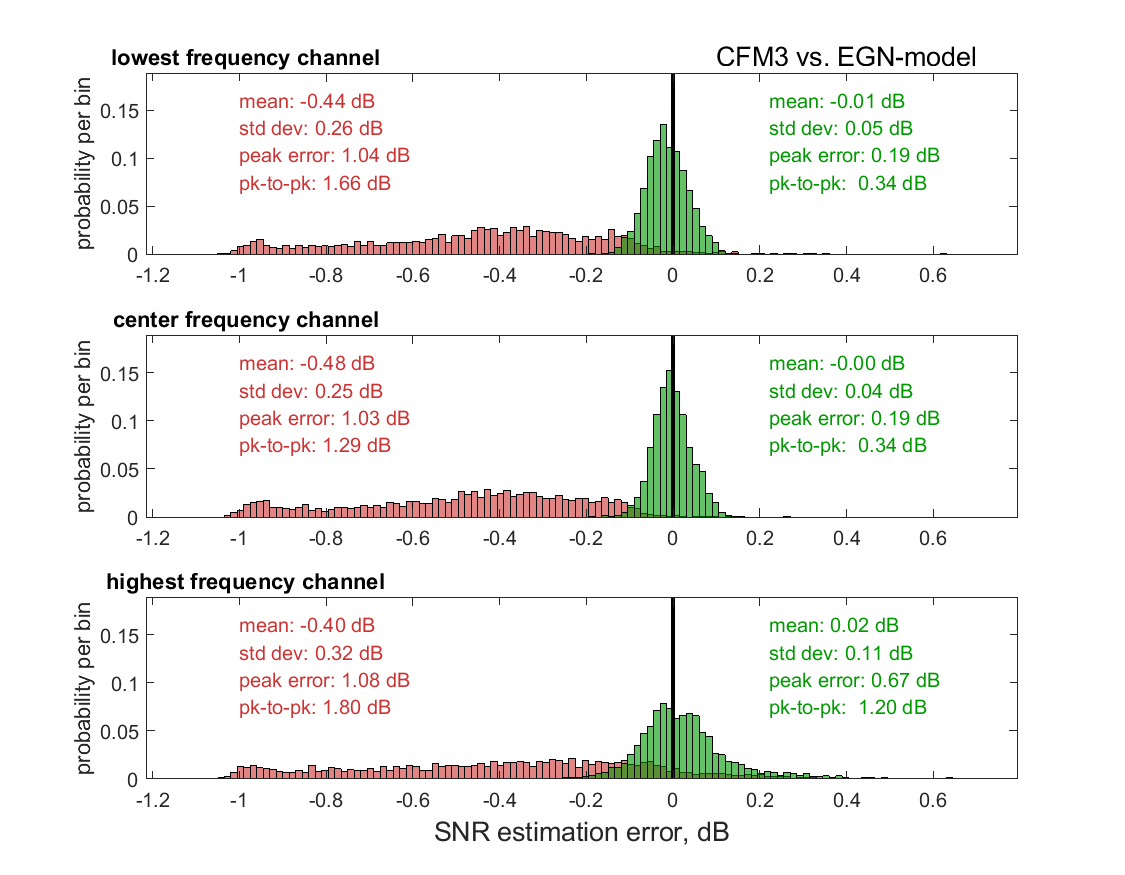}
\vspace{-0.4cm}
\caption{  Green bars: histograms of the SNR estimation error $\Delta^{{\rm{dB}}}_{_{\rm SNR}}$  between the closed-form model 3 (CFM3) and the EGN-model, as defined by Eq.~(\ref{eq:Delta}). The error is measured at maximum reach. Each  histogram was built by looking at about 2800 different systems. The red bars are CFM1 vs. EGN-model, same as Fig.~\ref{fig:CFM1-EGN}, shown here for comparison.
\label{fig:CFM3-EGN}}
\end{figure}

The test of CFM3 on the 8500 systems test-set yielded the results shown in Fig~\ref{fig:CFM3-EGN}. The improvement is quite substantial. All of the outliers for the lowest-frequency and the center-frequency CUTs completely disappeared, as shown by the values of the peak error (0.19~dB for both CUTs), essentially coinciding  with the visual footprint of the two histograms. Interestingly, even though still performing worse,  the high-frequency channel benefited from the extra term too, with all statistical indicators improving.

\begin{table}
\centering
\caption{Optimized values of the   parameters $a_{1}$-$a_{18}$ for the machine-learning factors of CFM3 \label{tab:CFM3}}
\label{TAB2}
\begin{tabular}{||c|c||c|c||}\hline
parameter & value & parameter & value   \\ \hline
$a_{1}$ & +9.1688e-1  & $a_{10}$ & -1.7810e0   \\ \hline
$a_{2}$ & -1.2188e0  & $a_{11}$& +9.8983e-1   \\\hline
$a_{3}$ & +1.1171e0 & $a_{12}$& -1.6009e+1   \\\hline
$a_{4}$ & -2.2566e+1 & $a_{13}$&  +1.0821e0  \\\hline
$a_{5}$ & +1.6405e0  & $a_{14}$& -1.1348e0   \\\hline
$a_{6}$ & -1.0075e0 & $a_{15}$&  +1.1140e-2  \\\hline
$a_{7}$ & +1.2266e+1 & $a_{16}$& +7.4397e+4   \\\hline
$a_{8}$ & +5.0115e-3  & $a_{17}$&  +1.3166e+3  \\\hline
$a_{9}$ & +8.0341e-1 & $a_{18}$& -2.0804e0    \\\hline
                
\end{tabular}
\end{table}

\section{CFM4: introducing roll-off in the\ machine-learning  factors}
\label{sect:roll-off}

\begin{figure}
\center
\includegraphics[width=\columnwidth]{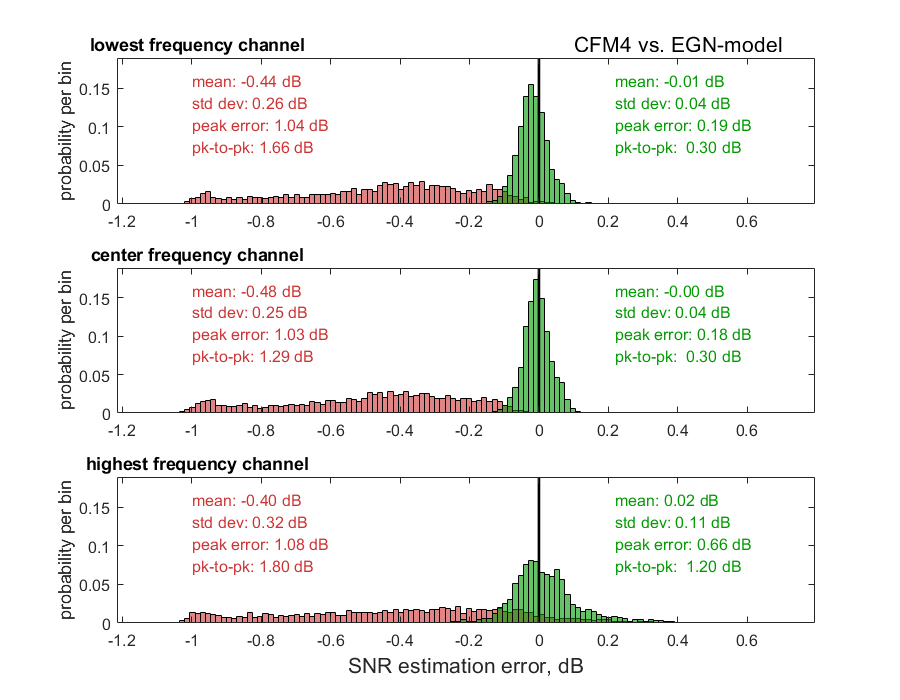}
\vspace{-0.4cm}
\caption{  Green bars: histograms of the SNR estimation error $\Delta^{{\rm{dB}}}_{_{\rm SNR}}$  between the closed-form model 4 (CFM4, accounting for WDM channels roll-off factors) and the EGN-model, as defined by Eq.~(\ref{eq:Delta}). The machine-learning factors definition used here is Eq.~(\ref{eq:roll-off}). The error is measured at maximum reach. Each  histogram was built by looking at about 2800 different systems. The red bars are CFM1 vs. EGN-model, same as Fig.~\ref{fig:CFM1-EGN}, shown here for comparison.\label{fig:CFM3ro-EGN}}
\end{figure}

In an effort to further improve the accuracy of CFM3, we introduced another system physical parameter in the machine-learning factors formulas, the WDM channels roll-off, as follows:
\begin{equation}
\begin{array}{l}
\rho _{{n_{{\rm{ch}}}}}^{(n)} = \left(1+ a_{19} \cdot r_{_{\rm CUT}}^{a_{20}} +a_{21} \cdot r_{n_{\rm ch}}^{a_{22}} \right)\cdot  \Big\{{a_1} + {a_2} \cdot {\Phi _{{n_{{\rm{ch}}}}}^{a_3}} + \\[0.1 cm] \hspace{1.7 cm} {a_4} \cdot {\Phi _{{n_{{\rm{ch}}}}}^{a_5}} \cdot    {\left( 1+ {a_6} \cdot  \left[\,{\left| {{\beta _{2,{\rm{acc}}}}\left( {n,{n_{{\rm{ch}}}}} \right)}  \right|  }+{a_{7}}\,\right]^{{a_{8}}} \right)}\Big\}
\\[0.1 cm]
\rho _{_{{\rm{CUT}}}}^{\left( n \right)} = \left(1+ a_{23} \cdot r_{_{\rm CUT}}^{a_{24}}\right) \cdot \Big\{{a_{9}} + {a_{10}} \cdot {\Phi _{_{\rm CUT}}^{a_{11}}} + {a_{12}} \cdot {\Phi _{_{\rm CUT}}^{a_{13}}} \cdot  \\[0.1 cm] \hspace{0.5 cm}{\left( 1+ {a_{14}} \cdot R_{_{\rm CUT}}^{a_{15}} + {a_{16}} \cdot  \left[\,{\left| {{\beta _{2,{\rm{acc}}}}\left( {n,{n_{_{\rm{CUT}}}}} \right)}  \right|  }+{a_{17}}\,\right]^{{a_{18}}} \right)}\Big\}
\\[0.5pt]
\label{eq:roll-off}
\end{array}
\end{equation}
where $r_{_{\rm CUT}}$ is the roll-off of the CUT and $r_{n_{\rm ch}}$ is the roll-off of the generic $n_{\rm ch}$-th channel.
The optimized values of the parameters $a_{1}\ldots a_{24}$ are shown in Table~\ref{tab:rho_ro}. The SNR error histograms obtained using the new machine-learning factors Eq.~(\ref{eq:roll-off}) are shown in Fig.~\ref{fig:CFM3ro-EGN}. We call this version of the model CFM4. 

Several of the histogram statistical indicators improve and no one gets worse, as a comparison between Fig.~\ref{fig:CFM3-EGN} and Fig.~\ref{fig:CFM3ro-EGN} shows. It therefore appears advantageous to use Eq.~(\ref{eq:roll-off})   rather than Eq.~(\ref{eq:rho2}), although the histograms of Fig.~\ref{fig:CFM3-EGN} are already so narrow that the gains shown in Fig.~\ref{fig:CFM3ro-EGN}  are modest. Yet, for the amount of added complexity, which is very limited, we think the gains are worthwhile.

In Appendix \ref{app:sens} an investigation of error sensitivity on the inclusion or exclusion of the other physical  parameters present in Eq.~(\ref{eq:roll-off}) is proposed, showing they need to be all kept in.

\begin{table}
\centering
\caption{Optimized values of the parameters $a_{1}$-$a_{24}$ for the machine-learning factors of CFM4 \label{tab:rho_ro}}
\label{TAB_with_roll_off}
\begin{tabular}{||c|c||c|c||}\hline
parameter & value & parameter & value   \\ \hline
$a_{1}$ & +1.0436e0  & $a_{13}$& +1.0229e0    \\\hline
$a_{2}$ & -1.1878e0  & $a_{14}$& -1.1440e0   \\\hline
$a_{3}$ & +1.0573e0 & $a_{15}$&  +1.1393e-2  \\\hline
$a_{4}$ & -1.8309e+1 & $a_{16}$& +3.8070e+5   \\\hline
$a_{5}$ & +1.6665e0  &  $a_{17}$&  +1.4785e+3  \\\hline
$a_{6}$ & -1.0020e0 &  $a_{18}$& -2.2593e0    \\\hline
$a_{7}$ & +9.0933e0 &  $a_{19}$& -6.7997e-1    \\\hline 
$a_{8}$ & +6.6420e-3  &  $a_{20}$& +2.0215e0    \\\hline
$a_{9}$ & +8.4481e-1 & $a_{21}$& -2.9781e-1    \\\hline 
$a_{10}$ & -1.8530e0  & $a_{22}$& +5.5130e-1    \\\hline                   
$a_{11}$& +9.4539e-1  &   $a_{23}$& -3.6718e-1    \\\hline 
$a_{12}$& -1.5421e+1 &   $a_{24}$& +1.1486e0    \\\hline 

\end{tabular}
\end{table}

\section{Full C-band split-step simulations}
\label{sect:split-step}
So far the CFMs have been compared with a benchmark which consists of the EGN-model. However
one could argue that at least one further independent and maybe more fundamental benchmark should be used to fully validate  the reliability of the CFMs. 

 We therefore decided to test CFM4 vs. full C-band split-step simulations.
Given the very high computational effort required,
only 300 different systems could be considered, 100 each with the lowest, center and highest frequency CUTs. They were randomly selected out of the 3150 PM-QAM systems in category (1). We believe that, though relatively small, this 300 system test-set is already large enough to provide an independent additional indication of whether CFM4 is a reliable NLI estimation tool or not.

Simulations were conducted according to standard, well-established techniques. The total number of `good symbols' per simulation was $2^{17}$. By `good symbols' we mean the remaining symbols after signal heads and tails were suitably discarded to ensure all transients extinction and to ensure that all-channels of the comb were simultaneously present while non-linearly interfering,  despite dispersion-induced channel slippage. The CUT receiver first compensated for nominal channel dispersion. A 2x2 complex LMS stage followed, which was initially operated in data-aided mode to achieve convergence and  was then `frozen'. Lasers were assumed ideal (zero linewidth).

\begin{figure}
\center
\includegraphics[width=\columnwidth]{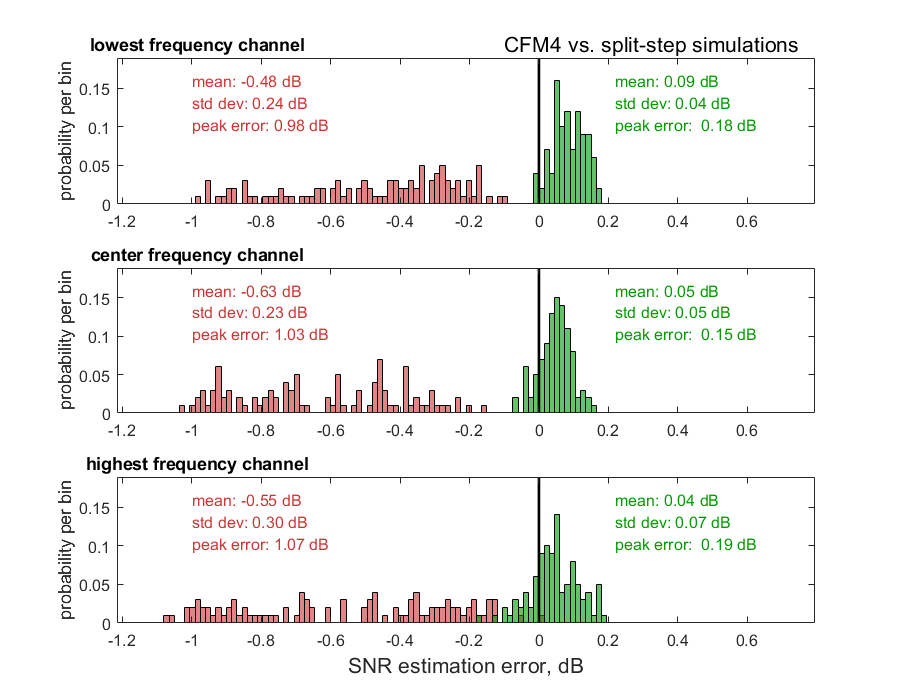}
\vspace{-0.4cm}
\caption{  Green bars: histograms of the SNR estimation error $\Delta^{{\rm{dB}}}_{_{\rm SNR}}$  between the closed-form model 4 (CFM4) and full C-band split-step simulations. The error is as defined by Eq.~(\ref{eq:Delta}) and measured at maximum reach. The simulations were performed using systems randomly chosen among the 3150 PM-QAM systems of category (1).  Each  histogram was drawn based on 100 different systems.  Red bars: CFM1 vs. simulations.
\label{fig:CFM3-SIM}}
\end{figure}

The results, shown in Fig.~\ref{fig:CFM3-SIM}, are very encouraging, in the sense that, with the inevitable statistical noise and uncertainty related to the smaller sample, the histograms appear to be quite similar to those of Fig.~\ref{fig:CFM3-EGN}. We believe this to be a very important result towards validating CFM4 and the overall closed-form modeling effort presented here. 

\section{Computational effort}
We used CFM4 to characterize all of the 8500 test-set systems. We did it on a laptop, using interpreted Matlab(TM) code. It took on average about 6 ms to calculate the SNR of \emph{all} WDM channels of a system. This is several orders of magnitude faster than using numerical integration of the EGN or even GN-model and it is seemingly compatible with real-time use.  


\section{Ongoing research}
The CFMs presented in this paper are currently being upgraded in two directions. On one hand, an effort is ongoing to validate the closed-form modeling of inter-channel stimulated Raman scattering (ISRS), in the version proposed in \cite{2018_arXiv_Poggiolini}. ISRS modeling is necessary for the correct appraisal of NLI in C+L band systems. On this topic substantial research is being done by other groups as well (\cite{2018_arXiv_Semrau}, \cite{2019_JLT_Semrau}), in view of the possible widespread upgrade of the existing network from C to C+L as currently installed capacity is gradually exhausted.   On the other hand, another effort is ongoing towards validating a CFM extension proposed in \cite{2019_arXiv_Ranjbar} which improves accuracy at low and near-zero dispersion. This effort aims at covering those links where legacy fibers are still found, whose dispersion is very low or even have an in-band dispersion-zero.  

\section{Conclusion}
 We derived and tested the accuracy of a closed-form approximate EGN-model formula over  8500  highly diversified C-band systems, of which 6250 were fully-loaded and 2250 partially-loaded. They used all combinations of   PM-QAM (from 4 to 256) formats, as well as PM-Gaussian, and three different fiber types (SMF and two types of NZ-DSF). Several other comb and link parameters were randomized as well. 

We then greatly improved the accuracy of the model by leveraging the large test-set, mimicking `big-data' approaches used in other contexts. We also added a new analytical term to deal with certain `corner cases' that the model could not handle well. In addition, we double-checked our validation by comparison with 300 full C-band split-step simulations. 

To the best of our knowledge, this is the first time such extensive testing and optimization procedures have been used for the modeling of the system impact of non-linear fiber propagation.

Away from pathological near-zero-dispersion situations, the final CFM (closed-form model) showed very good accuracy in reproducing the predictions of the full-fledged, numerically-integrated EGN-model, as well as the results of full-band split-step simulations, at a  comparatively negligible computational effort (on the order of milliseconds). 

We therefore believe the CFM proposed here  could potentially provide an effective and accurate tool to support real-time physical-layer-aware management and control of optical networks.

\appendix

\subsection{Launch Power Optimization Procedure}
\label{app:powopt}

Our goal in this paper was to carry out a thorough and challenging test of the CFM accuracy and reliability, over an extremely broad envelope of system configurations. We did not want to investigate specific system optimization strategies or techniques. 

With this premise, when faced with the problem of setting launch powers in the systems test-set, we did not aim for the absolute best system performance, which would require fine-tuning each channel power and would be quite difficult to achieve. Rather, we  aimed for CFM testing conditions that might be challenging but, at the same time, not completely unrealistic because of too high or too low launch powers. So we went for an approximate but manageable launch power optimization, with special attention paid to the CUT.





We started out from the initial assumption of each channel being launched  \emph{at the same PSD} as any other channel. If so, all channels have the same PSD as the CUT and we can write: \begin{equation}
\bar{G}_{{n_{{\rm{ch}}}}}^{(n)} = \bar{G}_{_{\rm{CUT}}}^{(n)}
\label{eq:launch0}
\end{equation} 
where, 
 following the same notation as in Eqs.~(\ref{eq:CFGN1})-(\ref{eq:beta}), $\bar{G}_{{n_{{\rm{ch}}}}}^{(n)}$ is the PSD of the $n_{\rm ch}$-th channel launched into the $n$-th span and 
 $\bar{G}_{_{\rm CUT}}^{(n)}$ is the PSD of the CUT into the $n$-th span.

We then modified Eq.~(\ref{eq:launch0}) by introducing a channel-by-channel\ randomization, as follows:
\begin{equation}
\bar{G}_{{n_{{\rm{ch}}}}}^{(n)} = \bar{G}_{_{\rm{CUT}}}^{(n)} \cdot {\xi _{{n_{{\rm{ch}}}}}}
\label{eq:launch}
\end{equation}
where the $\xi_{n_{\rm ch}}$'s are random variables uniformly distributed between 0.7 and 1.3, generated at launch and preserved for all spans.  For the CUT, however, we set the value of $\xi_{n_{\rm CUT}}$ to  1, so that   the CUT launch PSD was exactly $\bar{G}_{_{\rm CUT}}^{(n)}$. We performed the above randomization, because we wanted to make sure that the CFM was capable of accurately dealing with channel-by-channel non-uniformity, even of substantial extent.

We then considered the SNR for the CUT, at the end of the link, defined in Eq.~(\ref{eq:SNR}). 
By dividing both numerator and denominator by the CUT symbol rate, we transform powers into PSDs and we can re-write  Eq.~(\ref{eq:SNR}) as:
\begin{equation}
{\rm{SN}}{{\rm{R}}} = \frac{{{\bar{G}^{\rm Rx}_{_{\rm{CUT}}}}}}{{{{G}^{\rm Rx}_{_{\rm{ASE}}}} + {{G}^{\rm Rx}_{_{\rm{NLI}}}}}}
\label{eq:SNR2}
\end{equation}where all quantities are referred to the input of the receiver. We then remark that ${G}^{\rm Rx}_{_{\rm{NLI}}}$ in general depends on each channel launch power, into each span. Given Eq.~(\ref{eq:launch}), though, we can think of ${G}^{\rm Rx}_{_{\rm{NLI}}}$ as a function of just the  PSD of the CUT at the start of each span, that is a function of  $ \bar{G}_{_{\rm{CUT}}}^{(n)}$, with $n$ ranging from 1 to $N_{\rm span}$. This is because the CUT PSDs  contain all the information regarding the PSDs of all channels, at each span, through Eq.~(\ref{eq:launch}).

Eq.~(\ref{eq:SNR2}) could then be approximately maximized by using the  LOGO approach, and in particular Eqs.~(79), (80) in \cite{2014_JLT_Poggiolini}. Under the assumption of incoherent NLI accumulation, the LOGO approach ensures that global launch power optimization can be performed by optimizing launch power locally at each span.  
For the purpose of such span-by-span optimization, we used CFM1 to calculate NLI.
At the end of this procedure, we obtained the full set of the optimized PSDs of the CUT into each span. This would already provide a usable operating point for each system, since  all the information on launch PSDs for all channels is provided by just the  $ \bar{G}_{_{\rm{CUT}}}^{(n)}$, according to Eq.~(\ref{eq:launch}).

However, we decided to perform one further step. We point out that the information on all channels PSDs can be  provided by the CUT PSD \textit{into the first span only}  $ \bar{G}_{_{\rm{CUT}}}^{(1)}$ together with the set of all the EDFA gains ${\Gamma}^{(n)}$. In addition, it is possible to write the receiver SNR as follows:
\begin{equation}
{\rm{SN}}{{\rm{R}}} = \frac{{{\bar{G}^{\rm Rx}_{_{\rm{CUT}}}}}}{{{{G}^{\rm Rx}_{_{\rm{ASE}}}} + \left[{\bar{G}^{\rm (1)}_{_{\rm{CUT}}}}\right]^{3} \eta_{_{\rm NLI}}   }}
\label{eq:SNR3}
\end{equation}
where $\eta_{_{\rm NLI}}$ accounts for the strength of non-linearity in the overall link and depends on the EDFA gains ${\Gamma}^{(n)}$ but \emph{not} on $\bar{G}_{_{\rm{CUT}}}^{(1)}$. Under the assumption Eq.~(\ref{eq:launch}), this is true whatever model (GN, EGN, the CFMs) is used to calculate  $\eta_{_{\rm NLI}}$,  and in particular it does not depend on whether coherent or incoherent NLI accumulation is assumed.

We therefore calculated $\eta_{_{\rm NLI}}$ using the numerically-integrated EGN-model, while keeping the EDFA gains fixed in the first optimization step. We then re-optimized $\bar{G}_{_{\rm{CUT}}}^{(1)}$ to maximize Eq.~(\ref{eq:SNR3}). This new optimum value of  $\bar{G}_{_{\rm{CUT}}}^{(1)}$,   together with the  ${\Gamma}^{(n)}$ found after the first optimization step,  set all launch powers along the link through Eq.~(\ref{eq:launch}).

\subsection{Example of numerical investigation of the dependence on physical parameters of the machine-learning  factors}
\label{app:corr}

We provide here one example of the numerical investigations that we carried out to find which physical parameters affected the error between CFM1 and the EGN model. 

We defined the quantity $\delta^{(n)}_{_{\rm NLI}}$ as the increment, between the $n$-th span and the previous span, in the PSD of NLI affecting  the CUT:
\begin{equation}
    \delta^{(n)}=G^{(n)}_{_{\rm{NLI}}}\left( {{f_{_{\rm{CUT}}}}} \right)-G^{(n-1)}_{_{\rm{NLI}}}\left( {{f_{_{\rm{CUT}}}}} \right)
\end{equation}

\begin{figure}
\center
\includegraphics[width=\columnwidth]{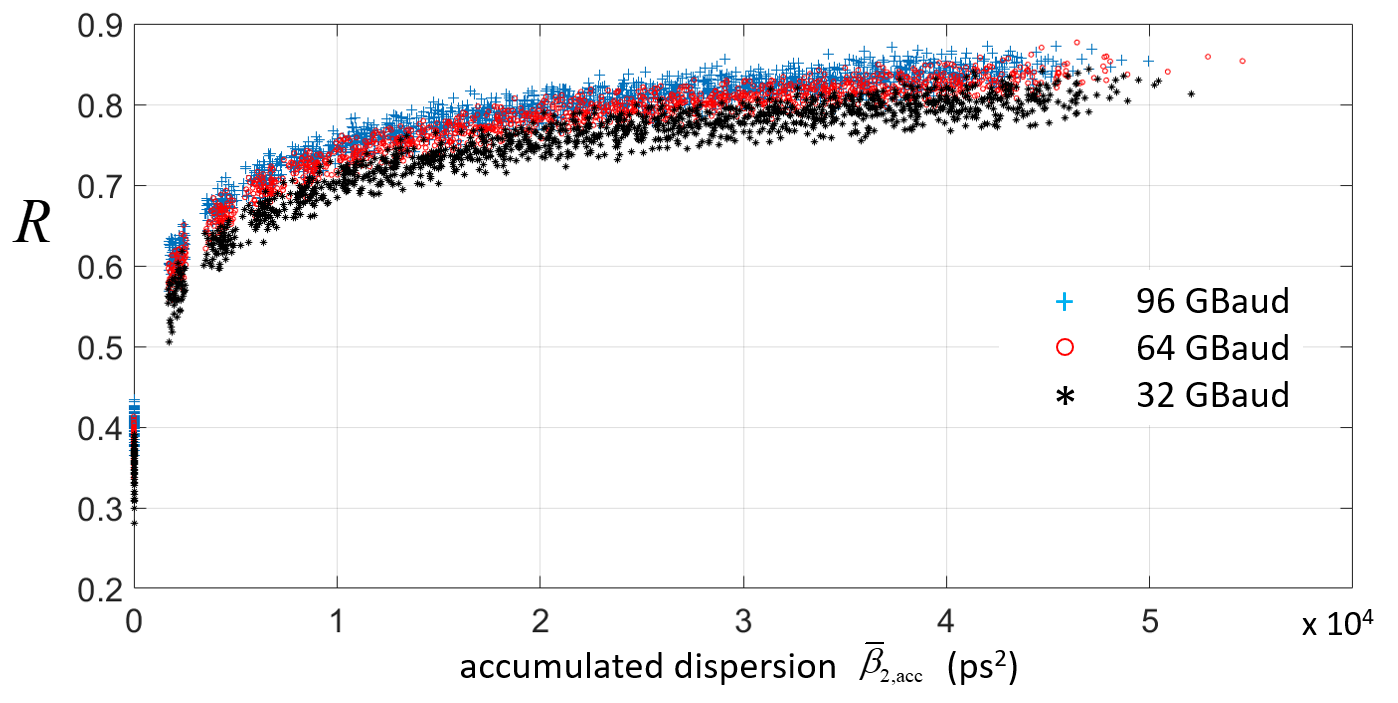}
\caption{
Plot of the quantity $R$ defined in Eq.~(\ref{eq:R}), representing the ratio between span-by-span increments of NLI as predicted by the EGN model and as predicted by the closed-form model 1 (CFM1). The abscissa is the accumulated dispersion at the start of each span. Spans were SMF  with   random length uniformly distributed between 80 and 120km. The tested channels were PM-QAM, from 16 to 256. The plot shows about 1500 values of the ratio  $R$. A value of 1 would mean same prediction between EGN and CFM1. Values lower than 1 mean that the EGN model predicts proportionally less NLI increment than CFM1.    
 \label{fig:evidence}
 }
\end{figure}

We estimated this quantity for \emph{each span} of 150 test systems (about 1500 data points) using category (1) WDM combs (PM-16QAM up to PM-256-QAM) and SMF links. The estimate was computed using both the EGN model ($\delta^{(n)}_{_{\rm EGN}}$) and CFM1 ($\delta^{(n)}_{_{\rm CFM1}}$). To compare the two estimates we then took their ratio:  
\begin{equation}R^{(n)}=\delta^{(n)}_{_{\rm EGN}}/\delta^{(n)}_{_{\rm CFM1}}
\label{eq:R}
\end{equation}
and plotted all the resulting 1500 ratios in Fig. \ref{fig:evidence}. 

As abscissa we used the absolute value of the accumulated dispersion for the CUT at the start of the $(n-1)$-th span. Therefore, the values of $R^{(1)}$ are lined up at the origin, i.e., their abscissa is ${{\bar \beta }_{{\rm{2}}{\rm{,acc}}}}$=0 because the accumulated dispersion at the start of the first span is obviously zero for all links. The values of  $R^{(2)}$ have abscissae that are spread $\pm 20$ \% about a mean value $|{{\bar \beta }_{{\rm{2}}{\rm{,acc}}}}|=0.213\cdot 10^{4}$ (ps$^{2}$), because SMF dispersion was $\beta_{2}$=$-21.3$ and the span lengths were \emph{uniformly distributed}  between 80 and 120 km. In turn the abscissae of  $R^{(3)}$ are spread  $\pm40$\%
with triangular distribution (the convolution of two identical uniform distributions). And so on, going to higher span numbers. 

A perfect coincidence between EGN model and CFM1 would imply $R^{(n)}=1$ for all $n$'s, i.e., all markers should have value 1 in the plot. Instead, substantial NLI increments overestimation occurs by CFM1, especially at the first spans, then gradually converging at farther spans or, equivalently, for higher accumulated dispersion. 

This trend does not come as a surprise, because CFM1 derives from the GN-model which is known to show such a behavior when compared to the EGN model. What is however quite interesting is that the dependence of the $R^{(n)}$'s on $|{{\bar \beta }_{{\rm{2}}{\rm{,acc}}}}|$ is remarkably smooth. Also, both evident and uniform is the dependence on the symbol rate, with different markers corresponding to different CUT symbol rates  making rather distinct `bands'. 

This evidence suggested to us that simple correction factors should be able to compensate for the errors, once properly trained through some `machine learning' process. This was done as shown in Sect.~\ref{sect:CFM2}, with good results.    

\subsection{Sensitivity of error vs.~physical parameters and comments on the analytical form of the machine-learning factors}
\label{app:sens}
The machine-learning factors shown in Eq.~(\ref{eq:roll-off}) depend on four physical parameters: 
\begin{enumerate}
\item the roll-off of each channel
\item the CUT symbol rate $R_{_{\rm CUT}}$
\item the modulation format of each channel, by means of the EGN model format-dependent coefficient $\Phi$, which relates to the second-moment of the transmitted constellation \cite{2014_OE_Carena}
\item the accumulated dispersion $\beta_{2,{\rm{acc}}}$ of each channel at each span 
\end{enumerate}

\begin{figure}
\center
\includegraphics[width=\columnwidth]{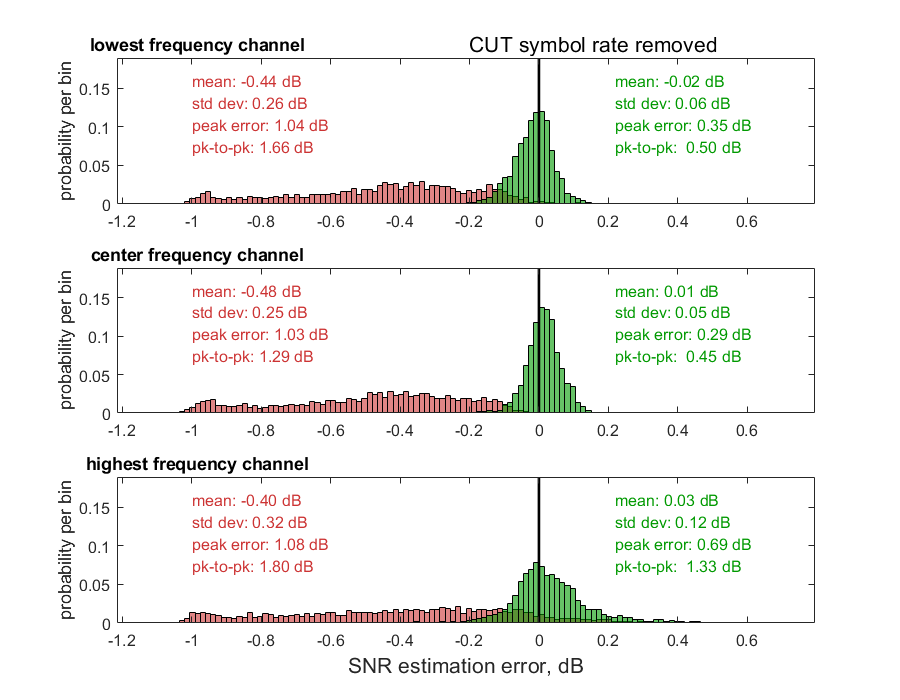}
\includegraphics[width=\columnwidth]{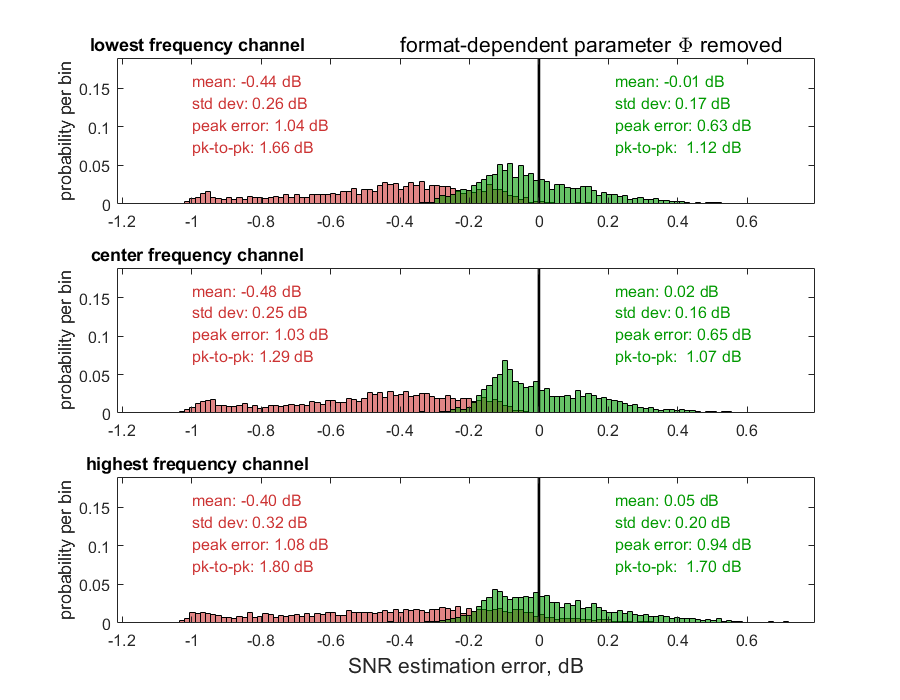}
\includegraphics[width=\columnwidth]{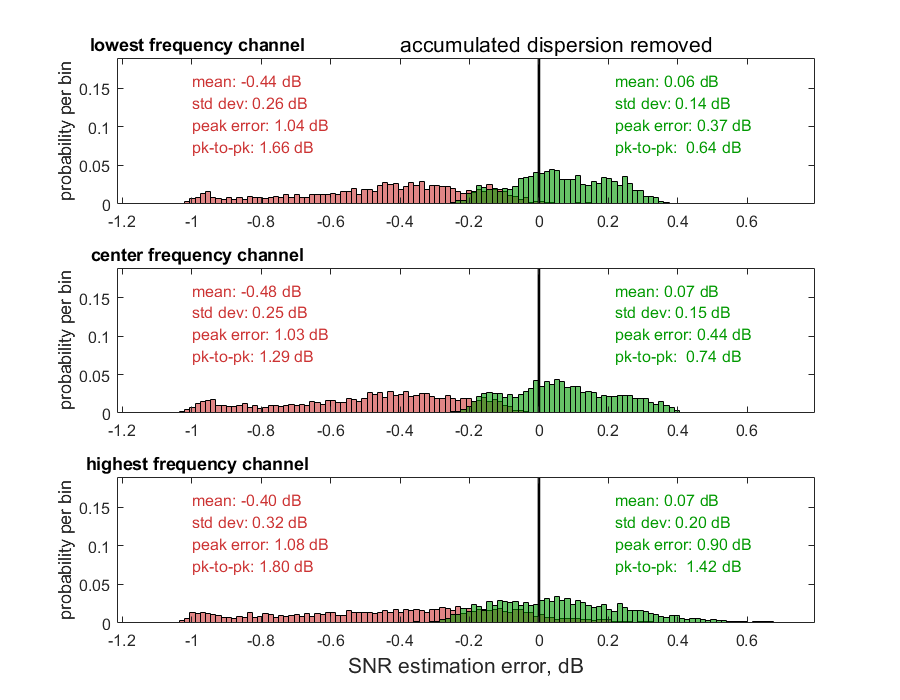}
\vspace{-0.4cm}
\caption{Green bars: histograms of the SNR estimation error $\Delta^{{\rm{dB}}}_{_{\rm SNR}}$  between the closed-form model 4 (CFM4) and the EGN-model, as defined by Eq.~(\ref{eq:Delta}). The machine-learning factors definition used here is Eq.~(\ref{eq:roll-off}). Top: CUT symbol rate removed from Eq.~(\ref{eq:roll-off}). Center: format-dependent parameter $\Phi$ removed from Eq.~(\ref{eq:roll-off}). Bottom: accumulated dispersion  $\beta_{2,{\rm{acc}}}$ removed from Eq.~(\ref{eq:roll-off}).
The red bars are CFM1 vs. EGN-model, same as Fig.~\ref{fig:CFM1-EGN}, shown here for comparison.\label{fig:comp}}
\end{figure}

We performed a `sensitivity' analysis of CFM4 error vs. these parameters. We did it by eliminating from  Eq.~(\ref{eq:roll-off}) in turn each one of the above parameters, while all others were left in. Note that when we removed a parameter we re-ran the best-fitting procedure on the free parameters $a_{n}$ so that each reduced version of the machine learning factors performed at its maximum potential.
In the following (Fig.~\ref{fig:comp}), we show the histograms of the error  $\Delta^{{\rm{dB}}}_{_{\rm SNR}}$, as defined by Eq.~(\ref{eq:Delta}), and the related statistical indicators,
when removing one of the parameters at a time. The benchmark for comparison consists of  the histograms and error statistical indicators vs. the EGN model, obtained when using the complete Eq.~(\ref{eq:roll-off}), that is CFM4, shown in Fig.~\ref{fig:CFM3ro-EGN}. We also collected the error statistical indicators, for the center channel  in the comb, in Table \ref{tab:stat-ind}.

The results clearly indicate that both the accumulated dispersion and the format-dependent parameter $\Phi$ are crucial for accurate modeling. Taking out either one of them very substantially degrades the performance of CFM4, with a tripling of variance and a doubling or tripling of peak and peak-to-peak errors. The impact of the CUT symbol rate being removed is small on mean and standard deviation, but the peak and peak-to-peak errors increase by 50\%. So it appears that such parameter is quite relevant too. As for channel roll-off, as discussed in Sect.~\ref{sect:roll-off}, its contribution is minor and could be neglected if simplifying the machine-learning factors was a priority. Also, note that the effect on mean value of the error of removing any parameter is always modest because the machine-learning formulas Eq.~(\ref{eq:roll-off}) have free parameters that absorb any fixed shift anyway.

To conclude, we propose a few remarks regarding the analytical form of the machine-learning factors  Eq.~(\ref{eq:roll-off}), which   may seem arbitrary  and perhaps in some parts redundant. 

One example is the  parameter $a_{7}$ whose presence may appear to be redundant, given that its value is about 9 and that it gets summed to  $|{{\bar \beta }_{{\rm{2}}{\rm{,acc}}}}|$, whose values are much greater than 9, even after just one span. However, as discussed in the previous appendix, $|{{\bar \beta }_{{\rm{2}}{\rm{,acc}}}}|$ is zero at the start of every link and such value is used by the CFMs to compute NLI over the first span. It is easy to verify that very different results would be found for this calculation if  $a_{7}$ was not there, causing substantial error. In particular, all systems with single-span reach, of which there are several instances in the 8500 system test-set,  would suffer from large NLI\ estimation errors.

Another example is provided by the overall factor, present in $\rho _{{n_{{\rm{ch}}}}}^{(n)} $:
\begin{equation}
\left( 1+ {a_6} \cdot  \left[\,{\left| {{\beta _{2,{\rm{acc}}}}\left( {n,{n_{{\rm{ch}}}}} \right)}  \right|  }+{a_{7}}\,\right]^{{a_{8}}} \right)\label{eq:ex2}
\end{equation}
Given that $a_{8}=6.6420\cdot 10^{-3}\approx 1/150$, this means taking the 150th root of the term: $\left[\,{\left| {{\beta _{2,{\rm{acc}}}}\left( {n,{n_{{\rm{ch}}}}} \right)}  \right|  }+{a_{7}}\,\right]$. This may raise the doubt that, essentially:
\[ \left[\,{\left| {{\beta _{2,{\rm{acc}}}}\left( {n,{n_{{\rm{ch}}}}} \right)} \right|  }+{a_{7}}\,\right]^{{a_{8}}} \approx 1 \]
If this was the case, then
\begin{equation}\left( 1+ {a_6} \cdot  \left[\,{\left| {{\beta _{2,{\rm{acc}}}}\left( {n,{n_{{\rm{ch}}}}} \right)}  \right|  }+{a_{7}}\,\right]^{{a_{8}}} \right)\approx \left( 1+ {a_6} \right)
\label{eq:dec}
\end{equation}
that is, a large part of the above formula would be redundant. In reality, an easy numerical check shows that the left and right-hand side of Eq.~(\ref{eq:dec}) differ by more than 10~dB over the almost entirety of the practical range of $|{{\bar \beta }_{{\rm{2}}{\rm{,acc}}}}|$, due to the near cancellation of the constant 1 with the remainder of the formula.

In conclusion, despite what may appear in some instances, all parts of the machine learning factors laws play a crucial role and deleting anyone  of them would impair accuracy. This is the case because we went through a thorough iterative `pruning' process where redundant parts were eliminated. While it is possible and indeed likely that simpler or more accurate laws can be devised, we think  Eq.~(\ref{eq:roll-off})  already achieves an effective combination of good performance and low overall complexity.

\begin{table}
\centering
\caption{Statistical error indicators for the center channel when removing different physical parameters from the machine-learning factors of CFM4 (all values in dB) }
\label{TAB_with_roll_off}
\begin{tabular}{||c||c|c|c|c||}\hline
 \bf removed parameter & mean & std dev & peak & peak-to-peak   \\ \hline
none & -0.00  & 0.04 & 0.18  & 0.30  \\\hline
channel roll-off & -0.00  & 0.05 & 0.19 & 0.34 \\\hline
CUT symbol rate & 0.01  & 0.05 & 0.29 & 0.45 \\\hline 
$\Phi$ (format) & 0.02 & 0.16 & 0.65 & 1.07 \\\hline
$ \beta_{2,{\rm{acc}}}$ & 0.07 & 0.15 & 0.44 & 0.74 \\\hline
\end{tabular}
\label{tab:stat-ind}
\end{table}

\end{document}